\documentclass[pra,amsmath,amssymb,amsfonts,nofootinbib]{revtex4}
\usepackage{amsmath,amsfonts,amssymb,amsthm}
\usepackage{bbm,graphicx,mathrsfs}
\usepackage[ansinew]{inputenc}

\providecommand{\ud}{\,\mathrm{d}}
\providecommand{\abs}[1]{\left\lvert#1\right\rvert}
\providecommand{\suchthat}{\,:\,}
\providecommand{\conj}[1]{\overline{#1}}
\providecommand{\hprod}[2]{\left\langle#1\,\vert\,#2\right\rangle}
\providecommand{\norm}[1]{\left\lVert#1\right\rVert}
\providecommand{\bra}[1]{\left\langle#1\right\vert}
\providecommand{\ket}[1]{\left\vert#1\right\rangle}
\providecommand{\avr}[1]{\left\langle#1\right\rangle}

\providecommand{\id}{\mathrm{id}}
\providecommand{\tr}{\mathrm{tr}}
\providecommand{\F}{\mathbb{F}}	

\providecommand{\Q}{\mathbb{Q}}
\providecommand{\R}{\mathbb{R}}
\providecommand{\C}{\mathbb{C}}
\providecommand{\quat}{\mathbb{H}}	
\providecommand{\M}{{\mathcal{M}}}
\providecommand{\1}{{\mathbbm{1}}}
\providecommand{\gr}[1]{\boldsymbol{#1}}
\providecommand{\be}{\begin{equation}}
\providecommand{\ee}{\end{equation}}
\providecommand{\bes}{\begin{equation*}}	
\providecommand{\ees}{\end{equation*}}
\providecommand{\G}{{\mathcal{G}}}
\providecommand{\HS}{{\mathfrak{H}}}
\renewcommand{\Re}{\mathrm{Re}}
\renewcommand{\Im}{\mathrm{Im}}

\newtheorem{definition}{Definition}
\newtheorem{theorem}[definition]{Theorem}
\newtheorem{proposition}[definition]{Proposition}
\newtheorem{lemma}[definition]{Lemma}
\newtheorem{corollary}[definition]{Corollary}

\begin{document}

\newenvironment{remark}{\vspace{1.5ex}\par\noindent{\it Remark}}%
    {\hspace*{\fill}$\Box$\vspace{1.5ex}\par}

\title{\sc \Large Unital Quantum Channels -- Convex Structure\\ and Revivals of Birkhoff's Theorem}

\author{Christian B. Mendl$^1$, Michael M. Wolf$^{1,2}$}
\affiliation{$^1$ Max-Planck-Institute for Quantum Optics, Garching, Germany\\
$^2$ Niels Bohr Institute, Copenhagen, Denmark}

\date{\today}

\begin{abstract}The set of doubly-stochastic quantum channels and its subset of mixtures of unitaries are investigated. We provide a detailed analysis of their structure together with computable criteria for the separation of the two sets. When applied to $O(d)$-covariant channels this leads to a complete characterization and reveals a remarkable feature: instances of channels which are not in the convex hull of unitaries can return to it when either taking finitely many copies of them or supplementing with a completely depolarizing channel. In these scenarios this implies that a channel whose noise initially resists any environment-assisted attempt of correction can become perfectly correctable.
\end{abstract}

\maketitle  \tableofcontents

\section{Introduction}

Quantum channels are the most general input-output relations which the framework of quantum mechanics allows for arbitrary inputs. Physically, they describe any transmission in space, e.g., through optical fibres, and/or evolution in time, as in in quantum memories, from a general open-systems point of view. Mathematically, they are characterized by linear, completely positive maps acting, in the Schr\"odinger picture, on density operators in a trace-preserving manner.

The present work investigates the particular class of quantum channels which leaves the maximally mixed (chaotic or infinite-temperature) state invariant. These channels are called \emph{unital} or \emph{doubly-stochastic} (referring to unital \emph{and} trace-preserving) and they appear naturally in contexts with an irreducible symmetry. Apart from their practical relevance, the interest in these channels has various origins: (i) they exhibit many special properties, e.g., regarding contractivity~\cite{PWPR06} or fixed points~\cite{AGG02} --- often allowing for a more geometric intuition, (ii) for small dimensions their additional constraint is strong enough to considerably simplify problems~\cite{K02}, and (iii) for sufficiently large dimensions problems on general channels can often be reduced to their unital counterparts~\cite{Fuk07,FW07,Ros08}.

The line of interest taken up by this article concerns the convex structure of the set of unital channels and, in particular, its relation to the subset of mixtures of unitary channels. This question was addressed and touched upon in~\cite{Tre86,KM87,LS93} where a crucial difference between the classical and the quantum case was realized: whereas, by Garrett Birkhoff's theorem~\cite{Bir46}, every doubly stochastic matrix (describing a classical channel) is a convex combination of reversible ones (i.e., permutations), not every doubly-stochastic quantum channel has to be a mixture of unitaries. The latter set became a stronger relevance when it was realized in Ref.\cite{GW03} that a quantum channel allows for perfect environment-assisted error correction if and only if it is a mixture of unitaries. Another remarkable step was made in Ref.\cite{SVW05} where evidence has been provided that asymptotically many copies of a unital channel might always be well approximated by a mixture of unitaries---a conjectured restoration of Birkhoff's theorem in the asymptotic limit.

An outline of the paper and a summary of its results:
\begin{itemize}
\item In Sec.\ref{sec:unital} we provide two characterizations of \emph{unital channels}: (i) as channels which are convex combinations of unitaries acting on Hilbert-Schmidt space, and (ii) as channels which are affine combinations of unitary channels. Moreover, we show that extreme points of the set of unital channels need not be extremal within the set of all channels.
\item In Sec.\ref{sec:unitary} computable criteria for the separation of unital channels from the set of mixtures of unitaries are provided and a respective negativity measure is introduced.
\item In Sec.\ref{sec:cov} we focus on covariant channels (in particular w.r.t. $O(d)$) and show how symmetry enables us to explicitly determine the above sets and to compute the negativity measure.
\item In Sec.\ref{sec:Birk} we apply the acquired tools in order show that families of covariant channels outside the convex hull of unitary channels fall back into this set when either taking several copies of them or supplementing with a completely depolarizing channel.
\end{itemize}

\section{Unital quantum channels}\label{sec:unital}

\subsection{Preliminaries}
We begin with introducing some notation and basic concepts. Throughout we will work in the Schr\"odinger picture and consider quantum channels $T$ with finite and equal input and output dimensions, i.e., $T:\mathcal{M}_d\rightarrow\mathcal{M}_d$ is a linear map on  $d\times d$ (density)matrices. Complete positivity enables a \emph{Kraus decomposition}
\be T(\rho)=\sum_i A_i \rho A_i^\dagger\label{eq:Kraus}, \quad \sum_i A_i^\dagger A_i = \1,\ee
where the second relation expresses the trace preserving property. A channel is called \emph{unital} if $T(\1)=\1$ and as we include the trace preserving property in the definition of a \emph{channel}, a unital channel is a \emph{doubly-stochastic} completely positive map.

It is often convenient to regard $\mathcal{M}_d$ as vector space which, when equipped with the inner product $\langle A,B\rangle:=\tr[A^\dagger B]$, forms the \emph{Hilbert-Schmidt}  Hilbert space $\mathfrak{H}_d$. Every channel is thus a linear map on this space  and has as such a respective matrix representation $\hat{T}\in\mathcal{M}_{d^2}\simeq\mathcal{B}(\HS)$.\footnote{$\mathcal{B}(X)$ denotes the space of \emph{bounded linear operators} on $X$.} We will occasionally use a (non-orthogonal) basis for $\HS$ which is obtained from embedded Pauli-matrices in the form
\be\label{eq:baseX}\begin{split}
\sigma_x^{jk} &:= \ket{j}\bra{k} + \ket{k}\bra{j} \quad \text{for all } j < k\\
\sigma_y^{jk} &:= -i \left(\ket{j}\bra{k} - \ket{k}\bra{j}\right) \quad \text{for all } j < k\\
\sigma_z^j &:= \ket{j}\bra{j} - \ket{j+1}\bra{j+1} \, \forall\,j = 1, \dots, d-1,
\end{split}\ee together with the identity matrix.

Another useful concept is the \emph{state-channel duality} introduced by Jamiolkowski~\cite{Jam72} which assigns a density operator $\rho_T\in\mathcal{M}_{d^2}$ to every channel $T$ via
\bes
\rho_T = (\id\otimes T)(\ket{\Omega}\bra{\Omega}),\quad \ket{\Omega}=\frac{1}{\sqrt{d}}\sum_{j=1}^d \ket{j,j}\ees
where  $\Omega$ is a \emph{maximally entangled state}.
The states $\rho_T$ corresponding to unital channels are exactly those with reduced density matrices
\be \tr_1\left[\rho_T\right] = \tr_2\left[\rho_T\right] = \1/d.\label{eq:RhoTpUnital}\ee
Note that due to the linearity of the correspondence the convex structure of channels is entirely reflected by the convex structure of the set of their dual states. Depending on what is more convenient we will switch back and forth between $T$ and $\rho_T$.

\subsection{Representations}\label{sec:represent}

In the remainder of this subsection we will prove the following characterization of unital channels:\newpage
\begin{theorem}[Characterization of unital channels]\label{thm:rep}
Let $T:\mathcal{M}_d\rightarrow\mathcal{M}_d$ be a quantum channel. Then the following are equivalent:\begin{enumerate}
\item $T$ is unital (i.e., doubly-stochastic),
\item $T$ is a convex combination of unitaries on $\HS_d$, i.e., $\hat{T}=\sum_\alpha p_\alpha W_\alpha$ with the $p$'s  being probabilities and each $W_\alpha\in\mathcal{M}_{d^2}$ unitary,
\item $T(\cdot)=\sum_i \lambda_i\,U_i\cdot U_i^\dagger$ is an affine combination of unitary channels, i.e., the $\lambda$'s are real and sum up to one and each $U_i\in\mathcal{M}_d$ is unitary.
\end{enumerate}
\end{theorem}
In order to see 1$\Leftrightarrow$2 we use a result from~\cite{PWPR06}: for any $p>1$ a positive trace-preserving map $T$ is a contraction in the sense\footnote{The norms are defined as $\norm{T}_{p\rightarrow q}=\sup_A \norm{T(A)}_q/\norm{A}_p$ with $\norm{A}_p=\big(\tr[(A^\dagger A)^{p/2}]\big)^{1/p}$.} of $\norm{T}_{p\rightarrow p}\le 1$ iff\footnote{As usual `iff' should be read `if and only if'.} $T$ is unital. In addition we have $\norm{T}_{2\rightarrow 2}=\big\lVert\hat{T}\big\rVert_\infty$ so that $T$ is unital iff $\hat{T}$ is a contraction with respect to the operator norm. The set of these contractions in turn is the convex hull of unitaries (as can be seen from the singular value decomposition) which completes 1$\Leftrightarrow$2.

As 3$\Rightarrow$1 is obvious it remains to show 1$\Rightarrow$3. To this end we introduce
\begin{align*}
\mathcal{X} &:= \left\{A \in \mathcal{M}_d \suchthat A = A^\dagger,\,\tr\,A = 0 \right\},\\
\mathcal{V} &:= \left\{A \mapsto U A U^\dagger \suchthat U \in \mathcal{M}_d\,\text{ unitary}\right\} \subset \mathcal{B}(\mathcal{X}).
\end{align*}
That is, $\mathcal{X}$ is a real linear subspace of $\HS$ containing all Hermitian operators orthogonal to $\1$ and $\mathcal{V}$ are the unitary conjugations on $\mathcal{X}$. Note that the real linear span of $\mathcal{V}$ is invariant under composition and that the set in Eq.~\eqref{eq:baseX} (without the identity) forms a basis of $\mathcal{X}$. The idea is now to show first how $\mathcal{B}(\mathcal{X})$ can be obtained from $\mathcal{V}$ and then to extend this to the claimed implication 1$\Rightarrow$3 in Thm.\ref{thm:rep}.

Denote the subspace of real linear combinations of vectors $\left\{x_1, \dots, x_n\right\}$ such that the coefficients sum to zero by
\bes
\mathrm{zerospan}_\R\left\{x_1, \dots, x_n\right\} := \left\{\sum_i \lambda_i\,x_i \suchthat \lambda_i \in \R,\, \sum_i \lambda_i = 0\right\}.
\ees
\begin{lemma}
For each basis vector $B\in \mathcal{X}$ in~\eqref{eq:baseX} there exists a $T \in \mathrm{zerospan}_\R \mathcal{V}$ which maps $B$ to itself and all other basis vectors to zero.
\end{lemma}
\begin{proof}
We explicitly construct such a $T$ w.l.o.g. for $\sigma_x^{12}$. Set
\begin{align*}
T_1(\rho) &:= \frac{1}{2} \left(\rho + U_1 \rho U_1^\dagger\right), \quad
U_1 := \left(\begin{array}{@{}c@{}c@{}}
\boxed{\1_2} &                \\
               & \boxed{-\1\,}\\
\end{array}\right),\\
T_2(\rho) &:= \frac{1}{2} \left(\rho - U_2 \rho U_2^\dagger\right) \in \mathrm{zerospan}_\R \mathcal{V},\quad
U_2 := \left(\begin{array}{@{}c@{}c@{}}
\boxed{\sigma_y} &                \\
                 & \boxed{\,\1\,}\\
\end{array}\right),
\end{align*}
Then for all $\alpha \in \R^4$ and $\sigma \equiv \left(\sigma_x, \sigma_y, \sigma_z, \1_2\right)$
\bes
A := \left(\begin{array}{c|c}
\alpha \cdot \sigma & B^*\\
\hline
B                   & C  \\
\end{array}\right)
\ \stackrel{T_1}{\mapsto} \
\left(\begin{array}{c|c}
\alpha \cdot \sigma & 0\\
\hline
0                   & C\\
\end{array}\right)\ \stackrel{T_2}{\mapsto} \
\left(\begin{array}{c|c}
\alpha_1 \sigma_x + \alpha_3 \sigma_z & 0\\
\hline
0                                     & 0\\
\end{array}\right).
\ees
In a similar vein we can finally map $\alpha_3$ to zero by a $T_3$, defined as $T_2$ only with $\sigma_y$ in $U_2$ replaced by $\sigma_z$. Then
$T := T_3 \circ T_2 \circ T_1$
is the desired operator which satisfies $T(A) = \alpha_1  \sigma_x^{12}$, and $T \in \mathrm{zerospan}_\R \mathcal{V}$ as $T_2 \in \mathrm{zerospan}_\R \mathcal{V}$. Clearly, the same type of construction works for all basis vectors in~\eqref{eq:baseX}.
\end{proof}
\begin{lemma}
For every pair of basis vectors $B_1, B_2\in \mathcal{X}$ in~\eqref{eq:baseX}, there is a $T \in \mathcal{V}$ such that $T(B_1) = B_2$.
\end{lemma}
\begin{proof}
As $B_1$ and $B_2$ are Hermitian, there are unitaries $U_1$ and $U_2$ such that $U_j^\dagger B_j U_j$ ($j = 1, 2$) are both diagonal. These can in turn be mapped onto each other by a permutation in $\mathcal{V}$ since they both have eigenvalues $(1, -1, 0, \dots, 0)$. Exploiting that $\mathcal{V}$ forms a group we can compose these steps to obtain $T(B_1) = B_2$.
\end{proof}
\begin{proposition}
$\mathcal{V}$ zero-spans all linear operators on $\mathcal{X}$, that is,
\bes \mathrm{zerospan}_\R \mathcal{V} = \mathcal{B}(\mathcal{X}).\ees
\label{prop:zerospan_unitary}
\end{proposition}
\begin{proof}
For any two basis vectors $B_1, B_2$ in~\eqref{eq:baseX}, by the above lemmas there is a $T \in \mathrm{zerospan}_\R \mathcal{V}$ which maps $B_1$ to $B_2$ and all other basis vectors to zero, so that a linear combination of these $T$'s generates any linear map on $\mathcal{X}$.
\end{proof}
This immediately implies 1$\Rightarrow$3 in Thm.\ref{thm:rep} as for every unital quantum channel $T$ we have that ${T-\id}\in\mathcal{B}(\mathcal{X})$ so that we can write $T(\rho) = \rho + \sum_i \lambda_i\,U_i \rho U_i^\dagger$ with the $\lambda$'s summing up to zero.

\subsection{Extreme points}

The set of all unital quantum channels on $\mathcal{M}_d$ is convex and compact. That is, every unital channel $T$ can be decomposed as \be T=\sum_i p_i T_i\ee where the $p$'s are probabilities and the $T_i$'s are \emph{extremal} unital channels, i.e., those which cannot be further decomposed in a non-trivial way. Despite considerable effort~\cite{Tre86,LS93,Rud04,Par05,AS07} not much is known about the explicit structure of these extreme points beyond $d=2$ (in which case they are all unitary conjugations~\cite{Tre86}). The small contribution of this subsection is to review the existing results and to apply them in order to show that channels which are extremal within the set of unital channels are not necessarily extremal within the convex set of all channels. To the best of our knowledge all known examples so far were extremal within both sets---although the  numerical results stated in~\cite{AS07} already indicate that this might not be generally true.
The main ingredient is the following theorem which is stated in~\cite{LS93} and based on~\cite{Cho75}.
\begin{theorem}[Extremal channels]
Consider a quantum channel with Kraus operators $\{A_i\}_{i=1,\ldots,N}$. It is an extreme point  within the convex set of quantum channels iff the set of matrices
\be\left\{A_k^\dagger A_l\right\}_{k,l=1 \dots N} \label{eq:extreme_channel}\ee
is linearly independent. Assume further that the channel is unital. Then it is extremal within the convex set of unital channels iff
\be \label{eq:extreme_unital_channel}
\left\{A_k^\dagger A_l \oplus A_l A_k^\dagger\right\}_{k,l=1 \dots N}\ee
is linearly independent.
\end{theorem}

We will exploit the fact that~\eqref{eq:extreme_channel} allows less linearly independent operators than~\eqref{eq:extreme_unital_channel}: while~\eqref{eq:extreme_channel} gives the simple bound $N\le d$, the set~\eqref{eq:extreme_unital_channel} yields $N\le \sqrt{2} d$.\footnote{In fact, in~\cite{Par05} it was shown that $N\le\sqrt{2d^2-1}$ which is, however, practically the same as $N\le \sqrt{2} d$ when applied to integer $N$.} For our example we choose dimension $d = 3$ and $N = 4$ linearly independent Kraus operators. The former ensures that there are non-trivial extreme points, and the latter already implies that~\eqref{eq:extreme_channel} can never be linearly independent as $N \not\le d$.
We start with an \emph{Ansatz} for the Jamiolkowski state of the sought channel of the form
\bes
\rho_T = (\id \otimes T)(\ket{\Omega}\bra{\Omega})\ =\ \sum_{i,j=1}^6 x_{ij} \ket{\psi_i} \bra{\psi_j},
\ees
where the $\left(\ket{\psi_i}\right)_i$ span the orthogonal complement of $\left(\ket{kk}\right)_k$, namely
\bes
\begin{split}
\ket{\psi_1} &= \frac{1}{\sqrt{2}\phantom{i}} \left(\ket{12}+\ket{21}\right), \quad
\ket{\psi_2} = \frac{1}{\sqrt{2}\phantom{i}} \left(\ket{13}+\ket{31}\right), \quad
\ket{\psi_3} = \frac{1}{\sqrt{2}\phantom{i}} \left(\ket{23}+\ket{32}\right),\\
\ket{\psi_4} &= \frac{1}{\sqrt{2} i} \left(\ket{12}-\ket{21}\right), \quad
\ket{\psi_5} = \frac{1}{\sqrt{2} i} \left(\ket{13}-\ket{31}\right), \quad
\ket{\psi_6} = \frac{1}{\sqrt{2} i} \left(\ket{23}-\ket{32}\right),\\
\end{split}
\ees
and the Hermitian matrix $X \equiv \left(x_{ij}\right)$ is given by
\be
X := \frac{1}{3} \begin{pmatrix}
\frac{1}{2}& 0& -i\,\mu_1& i\,\mu_3& i\,\mu_4& 0\\
0& \frac{1}{2}& -i\,\mu_1& -i\,\mu_4& -(2+i)\,\mu_3& 0\\
i\,\mu_1& i\,\mu_1& \frac{1}{2}& 0& 0& 2\,\mu_2 + i\,\mu_3\\
-i\,\mu_3& i\,\mu_4& 0& \frac{1}{2}& 0& -i\,\mu_1\\
-i\,\mu_4& (i-2)\,\mu_3& 0& 0& \frac{1}{2}& i\,\mu_1\\
0& 0& 2\,\mu_2 -i\,\mu_3& i\,\mu_1& -i\,\mu_1& \frac{1}{2}\\
\end{pmatrix}.
\label{eq:extremeX}\ee
The latter is chosen such that $\rho_T$ satisfies the conditions~\eqref{eq:RhoTpUnital} corresponding to a unital and trace-preserving map. It remains to choose algebraic numbers $\mu_1, \dots, \mu_4 \in \R$ such that $X$ is positive semidefinite with rank $N = 4$, and that at the same time~\eqref{eq:extreme_unital_channel} is linearly independent when plugging in the corresponding Kraus operators. A possible choice for such a set of parameters is provided in appendix~\ref{sec:extremeUnitalAppendix}.

\section{Mixtures of unitary channels}\label{sec:unitary}

This section deals with the class of unital channels which can be represented as
\be\label{eq:unitarym}
T(\rho) = \sum_{i=1}^N\; p_i U_i \rho U_i^\dagger,\quad U_i U_i^\dagger=\1, \ p_i > 0\ \forall\, i.\ee
The Jamiolkowski states corresponding to these mixtures of unitary conjugations are exactly the states which are convex combinations of maximally entangled states. The rank of the Jamiolkowski state $\rho_T$ gives a simple bound~\cite{Bus06} for the minimal $N$ as there exists always a decomposition with $N\le\left(\mathrm{rank}\,\rho_T\right)^2$. For $d=2$ we can achieve equality in the general lower bound $N\ge\mathrm{rank}\rho_T$ and, as mentioned before, every unital channel on $\mathcal{M}_2$ is a mixture of unitaries~\cite{Tre86}. For $d\ge 3$ the question whether a given unital channel allows for such a representation was investigated and  reformulated in~\cite{AS07} but a general operational way of deciding it remains to be found. The approach in the following subsection provides a class of easily computable necessary conditions which when applied to covariant channels will later be extended to necessary and sufficient criteria.

\subsection{Separation witnesses}\label{sec:sep_witness}

Since the set~\eqref{eq:unitarym} of mixtures of unitary channels is convex and compact, every unital channel which lies outside this set can be separated from it by a hyperplane --- a \emph{witness}. As this can most easily be expressed on the level of Jamiolkowski states we introduce the corresponding sets
\bes
\begin{split}
\mathcal{S} &:= \left\{\rho_T \suchthat \rho_T = (\id\otimes T)\left(\ket{\Omega}\bra{\Omega}\right), \ T: \mathcal{M}_d \to \mathcal{M}_d \text{ cp, tp, unital}\right\}\\
&\phantom{:}= \left\{\rho \in \mathcal{M}_{d^2} \suchthat \rho \ge 0, \ \tr_1 \rho = \tr_2 \rho = \1/d\right\},\\
\mathcal{U} &:= \mathrm{conv} \left\{\left(\1 \otimes U\right) \ket{\Omega}\bra{\Omega} \left(\1 \otimes U^\dagger\right) \suchthat U U^\dagger=\1\right\},
\end{split}
\ees
which we will, with some abuse of notation, occasionally also use for channels, i.e., we will write `$T\in\mathcal{S}$' meaning $\rho_T\in\mathcal{S}$.
The following shows that we may impose some structure on the witnesses --- they can be taken from the affine span of $\mathcal{U}$.
\begin{proposition}[Separation witnesses]
Let $\rho \in \mathcal{S}$ characterize a unital quantum channel. Then $\rho \in \mathcal{U}$, i.e., it is a mixture of maximally entangled states, iff
\bes \tr\left[W \rho\right] \ge 0 \ees
for all Hermitian operators $W \in \mathcal{M}_{d^2}$ which satisfy
\be\tr_1 W =  \tr_2 W = \1/d, \quad \tr\left[W \sigma\right] \ge 0 \ \forall\, \sigma \in \mathcal{U}.\label{eq:witnesscond}\ee
\end{proposition}
\begin{proof}
We have to show that if $\rho \notin \mathcal{U}$, then there exists such a $W$ with $\tr\left[W \rho\right] < 0$. First note that
\bes \mathcal{X} := \left\{A \in \mathcal{M}_{d^2} \suchthat A=A^\dagger, \ \tr_1 A = \tr_2 A = 0\right\}\ees
is a real linear space and $\mathcal{S} - \1/d^2 \subset \mathcal{X}$. Set $\tilde{\rho} := \rho - \1/d^2 \in \mathcal{X}$. Using the Hahn-Banach separation theorem~\cite[theorem~1.C in chapter~1]{Zeidler1995} we find a $\tilde{W} \in \mathcal{X}$ with
\bes
\tr\left[\tilde{W} \tilde{\rho}\right] < -1/d^2, \quad \text{but} \quad \tr\left[\tilde{W} \tilde{\sigma}\right] \ge -1/d^2 \ \forall\, \tilde{\sigma} \in \left(\mathcal{U}-\1/d^2\right).
\ees
Setting $W := \tilde{W} + \1/d^2$ yields the sought witness.
\end{proof}

To simplify matters we will in the following also consider Hermitian witnesses which do not fulfill the l.h.s. of~\eqref{eq:witnesscond} as long as the r.h.s. is satisfied. A class of this kind which turns out to be particularly useful are operators constructed from the \emph{flip} operator $\F: \ket{k,l} \mapsto \ket{l,k}$ in the form
\be
\label{eq:witAnsatz}
W = \left(\1\otimes B\right)\F\left(\1\otimes B^\dagger\right)+w(B)\,\1,\quad B\in\mathcal{M}_d,
\ee
where $w(B)\in\R$ is a constant depending on $B$ such that $W$ fulfills the r.h.s. in~\eqref{eq:witnesscond}. Before we determine this dependence let us note that replacing $(\1\otimes B)$ by $(A\otimes B)$ in Eq.~\eqref{eq:witAnsatz} won't lead to a more general class of witnesses since
$\left(A\otimes B\right)\F\left(A^\dagger\otimes B^\dagger\right)=\left(\1\otimes B A^\dagger\right)\F\left(\1\otimes A B^\dagger\right)$.

The sharpest constant $w(B)$ for which~\eqref{eq:witAnsatz} fulfills the witness condition $\tr[W\rho]\ge 0$ for all $\rho\in\mathcal{U}$ is obtained from
\begin{align*}
w(B)
&= -\min_U\tr\left[(\1\otimes B)\F\left(\1\otimes B^\dagger\right) (\1\otimes U)\ket{\Omega}\bra{\Omega}\left(\1\otimes U^\dagger\right)\right]\\
&= -\frac1d\min_U\tr\left[B^\dagger U B^T\conj{U}\right]\\
&= -\frac1d\min_A\big\{\tr\left[A\conj{A}\right]\suchthat \sigma(A)=\sigma(B)\big\},
\end{align*}
where $U$ is unitary, $A\in\mathcal{M}_d$ and $\sigma(A)$ denotes the singular values of $A$. We solve this matrix optimization problem in appendix~\ref{sec:AAbar} arriving at the following result.
\begin{theorem}[Tight witnesses]\label{thm:witness} For any $B\in\mathcal{M}_d$ with singular values $\sigma_1\ge\ldots\ge\sigma_d$ the operator in Eq.~\eqref{eq:witAnsatz} is a separation witness iff \be w(B)\ge \frac1d \left\{\begin{array}{ll}
2 \sum_{i=1}^{d/2} \sigma_{2i-1} \sigma_{2i},& d \text{ even}\\
2 \Big(\sum_{i=1}^{(d-1)/2} \sigma_{2i-1} \sigma_{2i}\Big) - \sigma_d^2,& d \text{ odd}.\\
\end{array}\right.\ee
\end{theorem}
Note  in particular that for $B=\1$ and $d$ odd we get $w\ge 1-2/d$ while for $d$ even $w\ge 1$. Hence, for even $d$ no channel is separated from $\mathcal{U}$ by such a  witness (since $\F+\1\ge 0$). However, we will see in Sec.\ref{sec:completepic} that for $d$ odd it becomes a powerful tool.

\subsection{A negativity measure}\label{sec:negativity}

There are several possible ways of quantifying the deviation of a channel $T\in\mathcal{S}\setminus\mathcal{U}$ from being a mixture of unitary channels: one may for instance follow~\cite{AS07}, use the \emph{entanglement of assistance}~\cite{SVW05,DFMSTU99} or the minimal distance to the set $\mathcal{U}$ w.r.t. some distance measure. The representation Thm.\ref{thm:rep} enables a very natural alternative approach---a \emph{base norm} (inspired by~\cite{VW02}). That is, the deviation is quantified by the smallest negative contribution when representing $T$ as an affine combination of terms in $\mathcal{U}$. More formally:
\begin{definition}[Negativity]
For all $\rho \in \mathcal{S}$ the \emph{base norm} associated with $\mathcal{U}$ is
\bes
\norm{\rho}_\mathcal{U} := \inf\left\{\alpha_p + \alpha_n \suchthat \rho = \alpha_p\,\sigma_p - \alpha_n\,\sigma_n,\ \alpha_{p,n} \ge 0,\ \sigma_{p,n} \in \mathcal{U}\right\},
\ees
and the corresponding \emph{negativity} is given by
\be
\mathcal{N}_\mathcal{U}(\rho) := \inf\left\{\alpha_n \suchthat \rho = \alpha_p\,\sigma_p - \alpha_n\,\sigma_n,\ \alpha_{p,n} \ge 0, \ \sigma_{p,n} \in \mathcal{U}\right\}.
\label{eq:negativity}
\ee
\end{definition}
For $\tr[\rho]=1$ the two are related via $\norm{\rho}_\mathcal{U}=1+2\mathcal{N_U}(\rho)$ and obviously $\mathcal{N_U}(\rho)=0$ iff $\rho\in\mathcal{U}$. The base norm behaves nicely under concatenation and convex combination. Writing $\norm{T}_\mathcal{U}:=\norm{\rho_T}_\mathcal{U}$ we get
\begin{proposition}
Let $T_i\in\mathcal{S}$ be a set of quantum channels and $p_i\ge 0$ probabilities then
\begin{align*}
\norm{\prod T_i}_\mathcal{U} &\le \prod \norm{T_i}_\mathcal{U},\quad \text{and}\\
\norm{\sum p_i\,T_i}_\mathcal{U} &\le \sum p_i \norm{T_i}_\mathcal{U}.
\end{align*}
\end{proposition}
Both can easily be proven from the definition. The latter can be interpreted as coming from  triangle inequality and homogeneity of the norm. Note also that the above norm is unitarily invariant in the sense of $\norm{T V}_\mathcal{U}=\norm{V T}_\mathcal{U}=\norm{T}_\mathcal{U}$ for every unitary conjugation $V$.

As always measures are easy to define but hard to compute. For covariant channels we will show the calculation in Sec.~\ref{sec:completepic}.

\section{Covariant channels}\label{sec:cov}

In order to arrive at more explicit results we need some help---coming in the form of symmetries imposed on the channels. Consider any subgroup $\G\subset U(d)$ with elements $g\in \G$ and two unitary representations $V_g, \tilde{V}_g$ on $\mathbb{C}^d$. We say that a channel $T:\mathcal{M}_d\rightarrow\mathcal{M}_d$ is $\G$-covariant w.r.t. these representations if for all $g\in \G$:
\be T\big(V_g\cdot V_g^\dagger\big)=\tilde{V}_gT(\cdot)\tilde{V}_g^\dagger.\label{eq:covT}\ee
In this sense the action of the channel `commutes with the symmetry'.
If $\tilde{V}$ is an irreducible representation then $T$ is unital as $T(\1)=\int dg\; T(V_gV_g^\dagger)=\int dg\; \tilde{V}_g T(\1)\tilde{V}_g^\dagger=\1$ by invoking Schur's Lemma (where $dg$ is the Haar measure). In order to express Eq.~\eqref{eq:covT} in terms of the Jamiolkowski state $\rho_T$ we introduce $G=\{\conj{V}_g\otimes \tilde{V}_g\}_{g\in\G}$ and its commutant $G'=\{X\in\mathcal{M}_{d^2} \suchthat \forall\ U_g\in G: [X,U_g]=0\}$. Covariance of the channel translates then simply to
\bes \rho_T\in G'.\ees
As we will see below most of the analysis can w.l.o.g. be restricted to this commutant which considerably simplifies matters as $\dim G'$ is for a sufficiently large symmetry group much smaller than $d^4$, the dimensionality we would have to deal with otherwise. The map
\bes \gr{P}(A) := \int dg\; U_g A U_g^\dagger\label{eq:twirl}\ees
defines a projection in $\mathcal{B}(\mathcal{M}_{d^2})$, often called \emph{twirl}, which maps every matrix $A$ into $G'$ and acts as the identity on $G'$.
Moreover, since $G'$ is an algebra it is spanned by a set of minimal projections $\{P_i\}$. These are orthogonal if $G'$ is abelian (which happens for large enough symmetry groups) so that every $X\in G'$ can be written as $X=\sum_i x_i P_i$ with $x_i=\tr[X P_i]/\tr[P_i]$. In this case we can easily determine\footnote{Here we have used that $\tr[\gr{P}(A)P_i]=\tr[A\gr{P}(P_i)]=\tr[AP_i]$.} \be \gr{P}(A)=\sum_i \frac{\tr[A P_i]}{\tr[P_i]} P_i.\ee
If $G'$ fails to be abelian a similar reasoning still applies---for a detailed exposition of these matters we refer to~\cite{VW02}. In order to see how covariance helps for our purposes let us denote the set of witnesses by $\mathcal{W}:=\{W=W^\dagger\suchthat \forall\sigma\in\mathcal{U}:\tr[W\sigma]\ge 0\}$.
\begin{proposition}[Reduction to the commutant]\label{prop:red}
Let $\rho\in\mathcal{S}\cap G'$ be the Jamiolkowski state corresponding to a covariant unital channel. Then $\rho\in\mathcal{U}$ iff $\tr[W\rho]\ge 0$ for all $W\in\mathcal{W}\cap G'$. Moreover,
\bes
\norm{\rho}_\mathcal{U} = \inf\left\{\alpha_p + \alpha_n \suchthat \rho = \alpha_p\,\sigma_p - \alpha_n\,\sigma_n,\ \alpha_{p,n} \ge 0,\ \sigma_{p,n} \in \mathcal{U}\cap G'\right\},
\ees
which equivalently holds for the negativity $\mathcal{N_U}$.
\end{proposition}
\begin{proof}
The crucial point for both parts is that $\sigma\in\mathcal{U}$ implies $\gr{P}(\sigma)\in\mathcal{U}$ which in turn means that $\gr{P}(W)\in\mathcal{W}$ for every $W\in\mathcal{W}$. Therefore due to $\tr[\rho \gr{P}(W)]=\tr[\gr{P}(\rho)W]=\tr[\rho W]$ the set $\mathcal{W}$ can w.l.o.g. be restricted to $G'$. Regarding the base norm we arrive at the stated result when starting with any optimal decomposition $\rho = \alpha_p\,\sigma_p - \alpha_n\,\sigma_n$ and applying the twirl to both sides of the equation.
\end{proof}
This suggests the program for the next subsections: fix a symmetry group, identify the commutant $G'$ and determine $\mathcal{U}$, $\norm{\cdot}_\mathcal{U}$ and $\mathcal{N_U}$ by exploiting the reduction to $G'$.

\subsection{$O(d)$ covariance}

\begin{figure}[!t]
\centering
\includegraphics[width=10cm]{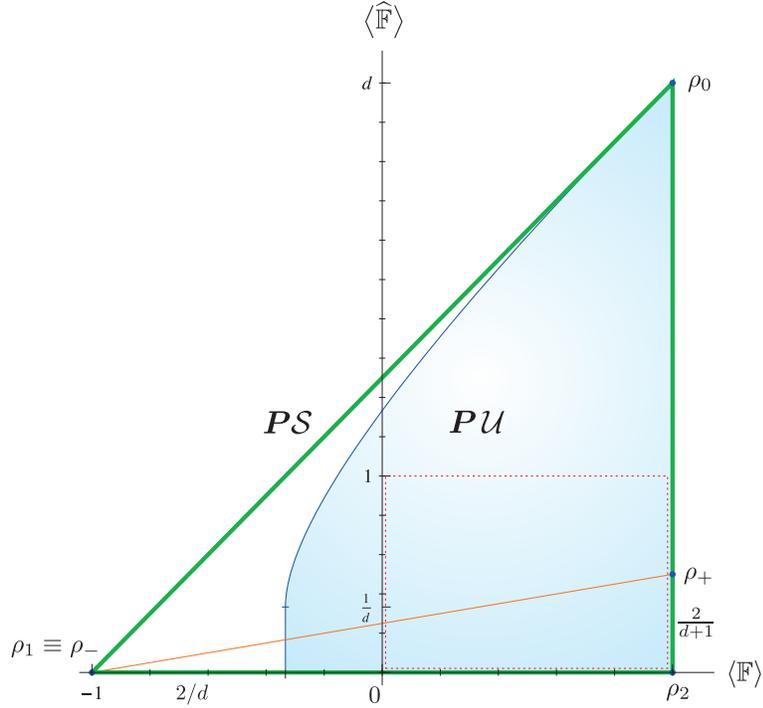}
\caption{The set of orthogonal covariant channels $\gr{P}\mathcal{S}$ in the Jamiolkowski representation (outer/green triangle) and the convex hull of unitary channels $\gr{P}\,\mathcal{U}$ (blue/shaded area), which is described in analytic terms by proposition~\ref{prop:traceUvsTraceUconjU1}. Note that the Werner-Holevo channel $\rho_-$ is ``furthest away'' from the unitaries. The orange line (from $\rho_-$ to $\rho_+$) depicts the $U(d)$ covariant channels, and the dotted unit-square corresponds to  entanglement-breaking channels. Compare with Ref.~\cite[Fig.~2]{VW01}.}
\label{fig:orth_covariant}
\end{figure}

The symmetry we will consider is the one of the real orthogonal group, i.e., $G = \left\{O \otimes O \suchthat O \in \mathcal{M}_{d} \text{ real orthogonal}\right\}$. The most prominent non-trivial example of a channel having this symmetry is
\be\label{eq:WH} T(\rho) = \left(\tr[\rho]\,\1-\rho^T\right)/(d-1),\ee
which (for $d=3$) gained some popularity as a steady source of counterexamples: for the multiplicativity of the output $p$-norm~\cite{WH02}, the additivity of the relative entropy of entanglement~\cite{VW01} and, most relevant in our context, the quantum analogue of Birkhoff's theorem~\cite{LS93}.
On the level of Jamiolkowski states we can make use of the analysis in~\cite{VW01} where the commutant $G'$ was shown to be abelian and spanned by
\bes
G' = \mathrm{span}\left\{\1, \F, \widehat{\F}\right\},
\ees
where $\widehat{\F}: = d\,\ket{\Omega}\bra{\Omega}$. From there the minimal projections can be identified as
\bes\begin{split}
P_0 &= \frac{1}{d}\,\widehat{\F} = \ket{\Omega}\bra{\Omega}\\
P_1 &= \frac{1}{2}\left(\1 - \F\right)\\
P_2 &= \frac{1}{2}\left(\1 + \F\right) - \frac{1}{d}\,\widehat{\F}\\
\end{split}\ees
where $(\1\pm\F)/2$ are the projections onto the symmetric and anti-symmetric subspace, respectively.
Consequently, every density operator in $G'$ is in the convex hull of the corresponding normalized density matrices $\rho_i = P_i/\tr[P_i]$ of which  $\rho_1$ corresponds to the \emph{Werner-Holevo channel} in~\eqref{eq:WH}, $\rho_0$ is the ideal channel and $\sum_i P_i/d^2$ corresponds  to the completely depolarizing channel $T(\rho)=\tr[\rho]\,\1/d$.  Clearly, all of them are unital, i.e., elements of $\mathcal{S}$. Every state $\rho \in G'$ is completely characterized by its "coordinates"
\bes\label{eq:orthcovariant_coords}
\left(\avr{\F}, \big\langle\widehat{\F}\big\rangle\right)_\rho \equiv \left(\tr\left[\rho \F\right], \tr\left[\rho\widehat{\F}\right]\right).
\ees
Especially for the extreme points $\rho_i$ we obtain (see Fig.~\ref{fig:orth_covariant})
\be\label{eq:covcoords_states}
\begin{tabular}{r|ccc}
state& $\rho_0$& $\rho_1$& $\rho_2$ $\vphantom{\bigl(}$\\
\hline
coords& $(1,d)$& $(-1,0)$& $(1,0)$ $\vphantom{\bigl(}$\\
\end{tabular}
\ee

\subsection{A complete picture}\label{sec:completepic}

We will now determine the subset $\mathcal{U}$ of mixtures of unitary channels within the set of $O(d)$-covariant channels. Following the above considerations this amounts to identifying the corresponding region in the two-dimensional parameter space
\be
\begin{split}
\mathcal{U}\cap G'
\cong&\left\{\left(\avr{\F}, \big\langle\widehat{\F}\big\rangle\right)_\rho \suchthat \rho \in \mathcal{U}\right\}\\
= \mathrm{conv}&\left\{\left(\avr{\F}, \big\langle\widehat{\F}\big\rangle\right)_U \suchthat U \text{ unitary}\right\},
\label{eq:coords_orthcov}
\end{split}
\ee
where the index $U$ stands for the expectation value w.r.t. $(\1\otimes U)\ket{\Omega}$ which parameterizes an extreme point within $\mathcal{U}$. A short calculation reveals that
\be\begin{split}
\avr{\F}_U
&\equiv \tr\left[\left(\1 \otimes U\right) \ket{\Omega}\bra{\Omega} \left(\1 \otimes U^\dagger\right) \F\right]
= \frac{1}{d}\,\tr\left[U \conj{U}\right],\\
\big\langle\widehat{\F}\big\rangle_U
&\equiv \tr\left[\left(\1 \otimes U\right) \ket{\Omega}\bra{\Omega} \left(\1 \otimes U^\dagger\right) \widehat{\F}\right]
= \frac{1}{d}\,\abs{\tr\,U}^2.\\
\end{split}\label{eq:eval_orthcov_unitary}\ee
The picture depends crucially on whether $d$ is even or odd.
\begin{theorem}[Even dimension]
If $d$ is even then $\mathcal{U}\cap G'=\mathcal{S}\cap G'$, i.e., every $O(d)$-covariant channel is a mixture of unitary channels.
\end{theorem}
\begin{proof}
It suffices to note that the expectation values~(\ref{eq:covcoords_states},\ref{eq:eval_orthcov_unitary}) with respect to $\rho_i$ and $U_i$ coincide for $U_0 = \1$, $U_1 = \mathrm{diag}\left(\sigma_y, \dots, \sigma_y\right)$ and $U_2 = \mathrm{diag}\left(\sigma_z, \dots, \sigma_z\right)$, just by plugging in~\eqref{eq:eval_orthcov_unitary}.
\end{proof}
So the interesting structure only emerges for $d$ odd (see Fig.~\ref{fig:orth_covariant} for $d = 3$), for which we need the following result proven in appendix~\ref{sec:maxTrUnitary}:

\begin{proposition}
Let $d \ge 1$ be odd. Then for all $x \in \left[-1+\frac{2}{d}, 1\right]$ there exists a unitary $U \in U(d)$ such that $ \tr\left[U \conj{U}\right]/d = x$, and
\be\begin{split}
&\max\left\{ \abs{\tr\,U}/d \suchthat U \in U(d),\, \tr\left[U \conj{U}\right]/d = x\right\}\\
&= \left[\frac{1}{2} \left(1 - \frac{1}{d}\right) \left(1 - \frac{2}{d} + x\right)\right]^{1/2} + \frac{1}{d}\ =:\ m(x).\\
\end{split}\label{eq:traceUvsTraceUconjU1}\ee
\label{prop:traceUvsTraceUconjU1}
\end{proposition}

\begin{theorem}[Odd dimension]\label{thm:oddd}
Let $d \ge 3$ be odd. Then the extreme points of the set~\eqref{eq:coords_orthcov} corresponding to mixtures of unitary channels are
\be \left(-1+2/d,0\right), \quad (1,0) \quad \text{and} \quad \left\{\left(x,d\big(m(x)\big)^2\right) \suchthat x \in \left[-1+2/d,1\right]\right\}.
\label{eq:extreme_orth_unitary}\ee
\end{theorem}
\begin{proof}
``$\eqref{eq:coords_orthcov} \subset \mathrm{conv}\eqref{eq:extreme_orth_unitary}$": For all unitary $U \in U(d)$,
\bes
\frac{1}{d}\,\tr\left[U \conj{U}\right] \in \left[-1 + 2/d, 1\right],
\ees
which follows from the fact that for any matrix $A$ the spectrum of $A \conj{A}$ is symmetric with respect to the real axis, the eigenvalues $\lambda, \conj{\lambda}$ have the same algebraic multiplicity, and the algebraic multiplicity of all negative eigenvalues of $A \conj{A}$ (if any) is even, see~\cite{FI06}. Together with Prop.~\ref{prop:traceUvsTraceUconjU1} we obtain the stated bounds on~\eqref{eq:coords_orthcov}.
\newline
``$\eqref{eq:extreme_orth_unitary} \subset \eqref{eq:coords_orthcov}$": Set
\bes Q_0 := \frac{1}{2}\begin{pmatrix}0&1-i&-1-i\\-1+i&-i&1\\1+i&1&i\end{pmatrix}\ees
and $\varphi := \exp\left(2 \pi i/3\right)$, then the coordinates~\eqref{eq:extreme_orth_unitary} are obtained by\\
$U_0 = \mathrm{diag}\left(\sigma_y,\dots,\sigma_y,Q_0\right)$, $U_1 = \mathrm{diag}\left(\sigma_z,\dots,\sigma_z,\varphi,\varphi^2,1\right)$ and the unitary matrices which solve the maximization problem~\eqref{eq:traceUvsTraceUconjU1} (explicitly given in appendix~\ref{sec:maxTrUnitary}).
\end{proof}

The fact that according to Prop.~\ref{prop:red}, we can restrict to decompositions within the two-dimensional parameter space, together with the explicit characterization of the set $\mathcal{U}\cap G'$ enables us now to compute the negativity $\mathcal{N_U}$, as follows.

\begin{figure}[!ht]
\centering
\includegraphics[width=10cm]{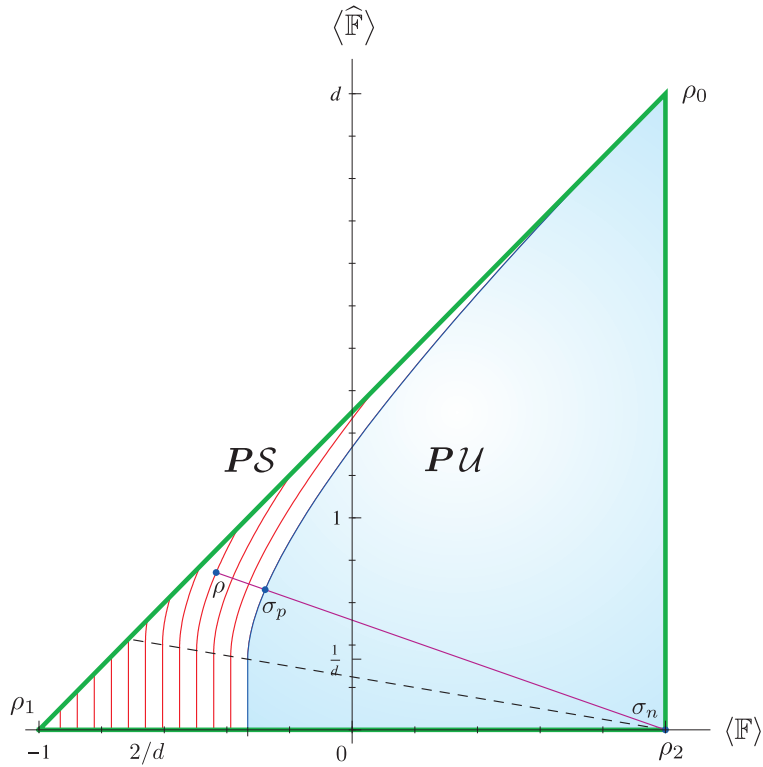}
\caption{Negativity as distance measure, exemplified by orthogonal covariant channels. $\mathcal{N_U}$ is constant along red lines, which are obtained -- in geometric terms -- by a uniform scaling of the unitary channel boundary about $\rho_2$ as origin. $\sigma_{p,n}\in\gr{P}\,\mathcal{U}$ are the optimal states in~\eqref{eq:negativity} given $\rho\in\gr{P}\mathcal{S}\setminus\gr{P}\,\mathcal{U}$.}
\label{fig:negativity}
\end{figure}

 We show first that in Eq.~\eqref{eq:negativity}, $\sigma_n = \rho_2$ always obtains the infimum, as illustrated in Fig.~\ref{fig:negativity}. Since $\mathcal{U}\cap G'$ is convex and closed, the optimal $\sigma_{p,n}$ in~\eqref{eq:negativity} are on the boundary of $\mathcal{U}\cap G'$, and $\sigma_n$ lies either on the segment joining $\rho_2$ and $\rho_0$ or the one joining $\rho_2$ and the covariant state with coordinates $\left(-1+2/d,0\right)$. We may w.l.o.g. assume the former, i.e., $\sigma_n = \lambda\,\rho_0 + (1-\lambda)\rho_2$ for a $\lambda \in [0,1]$. Considering decompositions $\rho = \alpha_p\,\sigma_p - \alpha_n\,\sigma_n$ with optimal $\sigma_p$ (depending on $\lambda$) given $\rho$ and $\sigma_n$ via $\lambda$, both $\alpha_{p,n}$ are already determined by the $x$-coordinates of $\rho$ and $\sigma_{p,n}$ due to $\alpha_p + \alpha_n = 1$. Note that the $x$-coordinates of $\rho$ and $\sigma_n$ remain fixed for all values of $\lambda$ whereas the $x$-coordinate of $\sigma_p$ is non-increasing as $\lambda$ decreases, and so is $\alpha_n$. That is, $\lambda = 0$ or equivalently $\sigma_n = \rho_2$ minimizes $\alpha_n$.

It follows that a uniform scaling of the boundary of $\mathcal{U}\cap G'$ by a factor $\left(1+\alpha_n\right)$ starting from $\rho_2$ as origin yields precisely the set of points with negativity $\alpha_n$.

We may write each $\rho\in\mathcal{S}\setminus\mathcal{U} \cap G'$ in terms of a convex combination of the $\rho_i$ listed in table~\ref{eq:covcoords_states}, that is, $\rho = \sum_i q_i\,\rho_i$ with $q_i \ge 0$ and $\sum_i q_i = 1$. From Fig.~\ref{fig:negativity} it is evident that $q_1 > 0$ for all $\rho \notin \mathcal{U}$. Set $q := q_0/q_1$ and distinguish the following two cases due to the particular shape of $\mathcal{U}\cap G'$.
\begin{itemize}
\item $q > \frac{1}{d(d-1)}$. This corresponds exactly to the area above the dashed line in Fig.~\ref{fig:negativity}. Applying the scaling $\left(1+\alpha_n\right)$ to the curve in theorem~\ref{thm:oddd}, an explicit calculation shows that
\begin{align*}
\mathcal{N_U}(\rho) &= \frac{1}{d-2}\left(d-1+\left(d+\frac{2}{d-2}\right)q-2\,\frac{d-1}{d-2}\sqrt{q^2+\frac{d-2}{d-1}\,q}\right) q_1 - 1,\\
\rho &= \sum_{i=1}^3 q_i\,\rho_i \in\mathcal{S}\setminus\mathcal{U}\ \text{ with } q_i \ge 0,\ \sum_i q_i = 1,\quad q := q_0/q_1.
\end{align*}
\item $q \le \frac{1}{d(d-1)}$. In this case the negativity does not  depend on $q$, and we get
\bes\mathcal{N_U}(\rho) = \frac{d}{d-1}\,q_1 - 1.\ees
\end{itemize}
In particular, $\mathcal{N_U}(\rho)$ is maximal exactly for the Werner-Holevo channel $\rho_-$, namely $\mathcal{N_U}\left(\rho_-\right) = 1/(d-1)$.

\section{Restoring Birkhoff's theorem}\label{sec:Birk}

Measures quantifying the deviation of a unital channel from being a mixture of unitary channels are known to be not additive (or multiplicative). That is, a naive extrapolation from the `distance' between a given $T\in\mathcal{S}\setminus\mathcal{U}$ and $\mathcal{U}$ typically leads to an overestimation of the respective quantity for $T^{\otimes n}$, i.e.,  several copies of the channel. This effect was studied in detail in the context of the \emph{entanglement of assistance}~\cite{DFMSTU99,SVW05}
\bes
E_a(\rho):=\sup\left\{\sum_i p_i S\big(\tr_1\Psi_i\big)\suchthat \rho=\sum_i p_i|\Psi_i\rangle\langle\Psi_i|\right\},\ees
where $S(\rho)=-\tr[\rho\log\rho]$ is the von Neumann entropy, and the supremum has to be taken over all convex decompositions of the given state $\rho\in\mathcal{B}(\C^d\otimes\C^d)$ into pure ones. As $S\left(\tr_1\Psi_i\right)\le \log d$ with equality iff $\Psi_i$ is a maximally entangled state we have that $E_a(\rho)\le\log d$ with equality iff $\rho\in\mathcal{U}$. It was shown in~\cite{SVW05} that
\bes
\forall\rho\in\mathcal{S}:\quad \lim_{n\rightarrow\infty} \frac1n E_a\left(\rho^{\otimes n}\right)=\log d,\ees
which suggests that the approximation of $\rho^{\otimes n}$ by an element of $\mathcal{U}$ improves as $n$ increases. This would mean a restoration of Birkhoff's theorem in the asymptotic limit.
Whether this statement is valid in general when formulated in terms of norm distances (either for channels or, supposedly weaker, for states) remains an open problem~\cite{problempage}.

In the following subsections we will prove it in the strongest possible sense for a class of $O(d)$-covariant channels. We will see that at least for these cases neither the asymptotic limit nor an approximation is required---a remarkable effect from the perspective of environment-assisted error correction (Sec.\ref{sec:envass}).

More specifically we will show that for a $T\not\in\mathcal{U}$ we find $T\otimes\tilde{T}\in\mathcal{U}$ when choosing
\be T: \M_d\to\M_d,\quad T(\rho)=\frac{1+\delta}d\;\tr[\rho]\,\1-\delta\,\rho^T,\quad d \text{ odd},\label{eq:Tex} \ee
with appropriate $\delta$ and either $\tilde{T}=T$ (Sec.\ref{sec:2copies}) or $\tilde{T}:\rho\in\mathcal{M}_D\mapsto \tr[\rho]\,\1\in\mathcal{M}_D$ completely depolarizing (Sec.\ref{sec:noisy}). The symmetry of the channels will help us in two stages: (i) we can use Thm.\ref{thm:oddd} which tells us that $T\not\in\mathcal{U}$ for $\delta>1/(d+1)$, and (ii) it circumvents having to find an explicit decomposition in terms of unitary channels for $T\otimes\tilde{T}$: if the convex hull of the relevant $G'$ expectation values of any set of unitary channels contains the ones of $T\otimes\tilde{T}$ then the twirling projection $\gr{P}$ does the rest of the job.

\subsection{Two copies of a channel}\label{sec:2copies}

\begin{figure}[!ht]
\centering
\includegraphics[width=10cm]{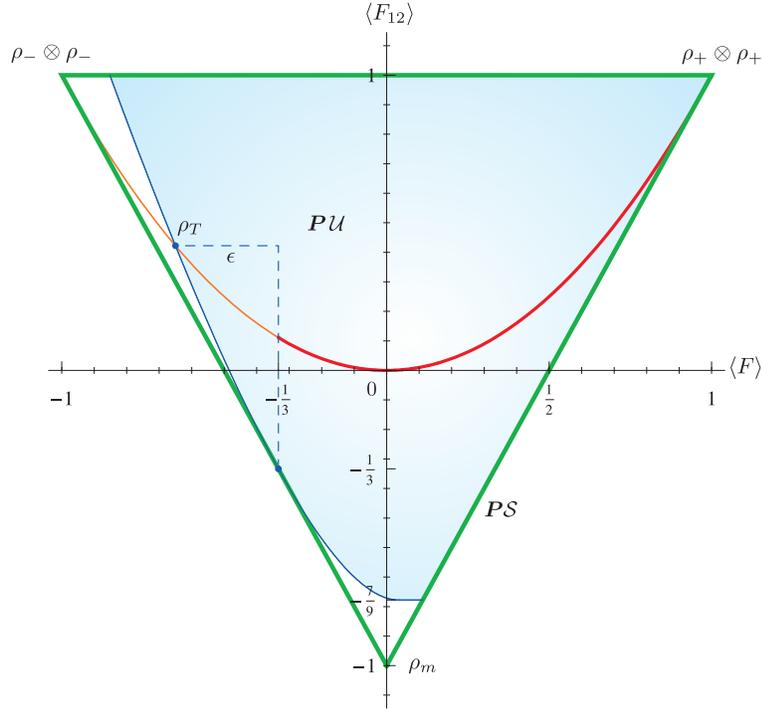}
\caption{Twofold tensor products of covariant channels for $d = 3$ (outer/green triangle). The mixtures of unitary channels correspond to the blue/shaded region, the orange parabola to tensor products of a channel with itself and the red part of that curve to the elements of $\mathcal{U}$ on the single-channel level. Compare with Fig.~9 in~\cite{VW01}.}
\label{fig:birkhoff}
\end{figure}

As usual we switch to the Jamiolkowski representation, where the family of channels in Eq.~\eqref{eq:Tex} becomes
\bes
\rho_T = \left(1 - \frac1d+\frac{\epsilon}2\right) \rho_- + \left(\frac1d-\frac{\epsilon}2\right) \rho_\text{+},\quad d\mbox{ odd}
\ees
where $\rho_\pm$ are the normalized projections onto the symmetric and anti-symmetric subspace, respectively. The parametrization is chosen such that $\rho_T\in\mathcal{S}\setminus\mathcal{U}$ for $\epsilon\in(0,2/d]$ since
\bes
\tr\left[\rho_T \F\right] = -1 + \frac{2}{d} - \epsilon,
\ees
and $\F+\left(1-2/d\right)\1$ is a tight separation witness according to Thms.\ref{thm:witness},\ref{thm:oddd}.
To exploit the full symmetry coming from $\tilde{T} = T$ we follow section V.B of~\cite{VW01} and increment the tensor product symmetry group\footnote{The aim is to use the full symmetry group. That is, we use $V\otimes V$, $V\in U(d)$ for $\rho_T$ which already allows us to discard $\hat\F$ from the commutant. We use further that the commutant of a tensor product is the tensor product of the commutants and that Prop.~\ref{prop:red} remains true when adding the additional flip operator.} $\left\{(U \otimes U\right) \otimes \left(V \otimes V)\right\}$ by a flip operator which interchanges the tensor factors in the product $T\otimes\tilde{T}$. This results in a larger symmetry group $G$, thus  yielding a smaller commutant $G'\subset\mathcal{M}_{d^4}$ which is spanned by $\1$ and
\bes F := \frac{1}{2}\left(\1\otimes\F + \F\otimes\1\right), \quad F_{12} := \F\otimes\F.\ees
That is, every state $\rho \in G'$ is now completely characterized by the expectation values/coordinates
\bes \left(\avr{F},\avr{F_{12}}\right)_\rho \equiv \left(\tr\left[\rho\,F\right], \tr\left[\rho\,F_{12}\right]\right).\ees
Especially for any unitary channel described by $U\in \mathcal{M}_{d^2}$, setting $U_\mathfrak{s} := \frac{1}{2}\left(U+U^T\right)$ and denoting partial transposes by $^{T_i}$ gives
\be\label{eq:uig}
\begin{split}
\avr{F_{12}}_U
&= \tr\left[F_{12} \left(\1 \otimes U\right)\ket{\Omega} \bra{\Omega}\left(\1 \otimes U^\dagger\right)\right] = \frac{1}{d^2}\,\tr\left[U\,\conj{U}\right],\\
\avr{F}_U
&= \tr\left[F \left(\1 \otimes U\right)\ket{\Omega}\bra{\Omega}\left(\1 \otimes U^\dagger\right)\right]\\
&= \frac{1}{2\,d^2}\,\tr\left[U\left(\conj{U}^{T_1}+\conj{U}^{T_2}\right)\right]\\
&= \frac{1}{d^2}\,\tr\left[U\,\conj{U_\mathfrak{s}}^{T_2}\right] = \frac{1}{d^2}\,\tr\left[U_\mathfrak{s}\,\conj{U_\mathfrak{s}}^{T_2}\right].\\
\end{split}\ee
The last equation uses the fact that $\conj{U_\mathfrak{s}}^{T_2}$ is again symmetric. The ranges of the expectation values in Eq.~\eqref{eq:uig} are studied in appendix~\ref{sec:minTrPTSymU}. In particular for $d=3$ we provide an explicit construction for the coordinates
\be \left(\avr{F},\avr{F_{12}}\right)_\vartheta = \frac{1}{9}\left(-\frac{8}{3}\left(\cos\vartheta + 1\right)^2+3,\,16\cos^2\vartheta-7\right), \quad \vartheta \in \left[0,\pi/2\right]
\label{eq:coords_theta}\ee
corresponding to convex combinations of unitary channels.
Now matching the coordinates of $T \otimes T$,
\bes
\left(\avr{F},\avr{F_{12}}\right)_T = \left(\tr\left[\rho_T \F\right], \tr\left[\rho_T \F\right]^2\right) = \left(x,x^2\right), \quad x := -1+\frac{2}{d}-\epsilon
\ees
with~\eqref{eq:coords_theta} yields
\bes
\epsilon = \frac{2}{3}\left(4-3\,\sqrt{2}-\sqrt{3}+\sqrt{6}\right) \approx \frac{1}{\sqrt{10}}
\ees
as shown in Fig.~\ref{fig:birkhoff}. The blue area corresponds to convex combinations of unitary channels\footnote{A proof that~\eqref{eq:coords_theta} really solves the minimization problem~\eqref{eq:minTrPTSymU} as acclaimed is still outstanding, but strongly supported by numerical tests. In any case, the set of convex combinations of unitary channels is \emph{at least} as big as the blue area.}, i.e., elements of $\mathcal{U}$, the orange curve to coordinates of single-channel tensor products and the red part of this curve to the elements of $\mathcal{U}$ on the single-channel level. The state at the lower corner is
\bes
\rho_m = \frac{1}{2}\left(\rho_-\otimes\rho_+ + \rho_+\otimes\rho_-\right).
\ees
As can be seen from direct inspection, each point on the curve~\eqref{eq:coords_theta} is an extreme point of the blue area. The remaining extreme points $(1,1)$ and $\left(\frac{1}{9},-\frac{7}{9}\right)$ are realized by the unitaries $\1$ and $\left(\begin{smallmatrix}&-1\\1&\end{smallmatrix}\right)^{\otimes 4} \oplus 1$, respectively.

\subsection{Help from a noisy friend}\label{sec:noisy}

Instead of adding a second copy of the channel, we will now supplement it by a completely depolarizing $\tilde{T}$. The Jamiolkowski representation of the completely depolarizing channel $\tilde{T}: \mathcal{B}\left(\mathcal{K}_1\right) \to \mathcal{B}\left(\mathcal{K}_2\right), \mathcal{K}_i=\mathbb{C}^D$ is
\bes
\rho_{\tilde{T}} \equiv \big(\id\otimes\tilde{T}\big)(\ket{\Omega}\bra{\Omega}) = \1/D^2.
\ees
Let $H$ be the corresponding symmetry group of all local unitaries $V_1 \otimes V_2$ with $V_i \in \mathcal{B}\left(\mathcal{K}_i\right)$, then $H' = \mathrm{span}\{\1\}$. Using again that
\bes
\left(G \otimes H\right)' = G' \otimes H'
\ees
(see example 7 in section II.D of~\cite{VW01}) we get that every element of the commutant is completely characterized by the expectation value of $Y = \F_{\mathcal{H}_1\otimes\mathcal{H}_2}\otimes\1_{\mathcal{K}_1\otimes\mathcal{K}_2}$, yielding
\be
\tr\left[\left(\rho_T\otimes\rho_{\tilde{T}}\right) Y\right] = \tr\left[\rho_T \F\right] = -1 + \frac{2}{d} - \epsilon.
\label{WitnessTensor}
\ee
Since $-\1\le Y\le\1$  every normalized state $\rho$ satisfies $\tr\left[\rho Y\right] \in \left[-1, 1\right]$.
In order to obtain the subinterval of $\left[-1, 1\right]$ covered by convex combinations of unitary channels, we have to calculate $\avr{Y}_U$ for unitary $U:\mathcal{H}_1\otimes\mathcal{K}_1 \to \mathcal{H}_2\otimes\mathcal{K}_2$,
\be
\tr\left[\left(\1 \otimes U\right)\ket{\Omega}\bra{\Omega}\left(\1 \otimes U^\dagger\right) Y\right] = \frac{1}{d\,D} \tr\left[U\,\conj{U}^{T_2}\right].
\label{CoordDepU}
\ee
As $U = \1$ reaches the upper bound $1$, the hard part is the lower bound which is treated in appendix~\ref{sec:minTrPTUnitary}. The results suggest that for $D = 2$, Eq.~\eqref{CoordDepU} gives
\be\label{eq:conjint}
\left[-1+\frac{2}{d^2},1\right]
\ee
for the range in which $T\otimes\tilde{T}\in\mathcal{U}$ while, recall, $T\in\mathcal{U}$ only within $[-1+2/d,1]$. The interval~\eqref{eq:conjint} can be related to the conjectured existence of a certain quaternion matrix, which we construct explicitly for $d = 3$ and $d = 5$. This means that in this case~\eqref{CoordDepU} covers this range \emph{at least}. In particular, for $\epsilon\le 2(d-1)/d^2$, the expectation value~\eqref{WitnessTensor} lies within this interval such that $\rho_T\otimes\rho_{\tilde{T}}$ becomes then indeed a convex combination of maximally entangled states.

For higher values of $D$, we reproduce table~\ref{tab:minTrPTUnitary} from appendix~\ref{sec:minTrPTUnitary}.
\begin{table}[!ht]
\centering
\begin{tabular}{r||rcccc}
$d$& $D = 1$& 2& 3& 4& 5\\
\hline
1& 1& 1& 1& 1& 1 $\vphantom{\Bigl(}$\\
3& $-\frac{1}{3}$& $-\frac{7}{9}$& $-\frac{23}{27}$& $-\frac{4}{5}$& $-\frac{37}{45}$ $\vphantom{\Bigl(}$\\
5& $-\frac{3}{5}$& $-\frac{23}{25}$& $-0.929\dots$ & ${-1}$& $-0.976\dots$ $\vphantom{\Bigl(}$\\
7& $-\frac{5}{7}$& $-\frac{47}{49}$& & & $\vphantom{\Bigl(}$\\
\end{tabular}
\caption{Numerical minimization of~\eqref{CoordDepU}. Note that the table is asymmetric w.r.t $d \leftrightarrow D$, and  that the lower bound $-1$ can apparently be obtained for $d=5$ and $D=4$, that is, each channel-tensor product $T\otimes\tilde{T}$ (with $\tilde{T}$ the completely depolarizing channel) becomes then a convex combination of unitaries.}
\label{tab:minTrPTUnitary1}
\end{table}

\subsection{Environment-assisted error correction}\label{sec:envass}

\begin{figure}[!ht]
\centering
\includegraphics[width=10cm]{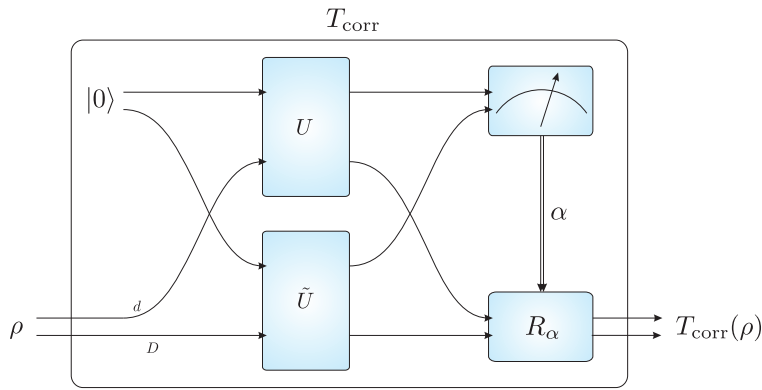}
\caption{The correction scheme as in~\cite{GW03}, applied to the simultaneous usage of two noisy channels $T: \mathcal{B}(\mathcal{H}_1) \to \mathcal{B}(\mathcal{H}_2)$ and $\tilde{T}: \mathcal{B}(\mathcal{K}_1) \to \mathcal{B}(\mathcal{K}_2)$. The channels are represented by unitary couplings $U$ and $\tilde{U}$ to an environment which is initially in a pure product state. The classical result $\alpha$ of the measurement on the global environment is used by the receiver who chooses the recovery operation $R_\alpha$ (again a quantum channel) accordingly. As discussed in the text, $T\otimes\tilde{T}$ can become perfectly correctable (i.e., $T_{\mathrm{corr}}=\id$) although neither $T$ nor $\tilde{T}$ is so.}
\label{fig:corr_scheme_tensor}
\end{figure}

The above results become especially remarkable from the point of view of environment-assisted error correction---a concept introduced in~\cite{GW03}. There it was studied which channels allow complete correction, given a suitable feedback of classical information from the environment (see Fig.~\ref{fig:corr_scheme_tensor}). The class of perfectly correctable channels was identified with the set of convex combinations of unitary channels. In this way the above observations yield examples of channels which are not perfectly correctable on their own but become so when either taking several copies, or supplementing with a completely depolarizing channel.

\section{Discussion}

The presented investigation of the set of unital quantum channels is to a large extent based on and inspired by methods and ideas from entanglement theory. The tools acquired in this context could be directly applied to the Jamiolkowski representation of the channel. This approach as such leads to questions about further analogies between the two fields. It would in particular be interesting whether a useful counterpart to \emph{positive maps}, i.e., powerful non-linear criteria can be found for the separation of the set of mixtures of unitary channles\footnote{Here it might be helpful to replace positivity by some form of contractivity.}.

Clearly, the asymptotic Birkhoff conjecture\cite{problempage} remains an important open problem for which the present work might be regarded as supporting evidence as it provides the first class of examples for which there is a rigorous proof. In this context it might be interesting to investigate $T\otimes\tilde{T}_D$ with $\tilde{T}_D$ a $D$-dimensional maximally depolarizing channel, as studied in Sec.\ref{sec:noisy}. Is there a dense subset of unital channels such that a finite $D$ makes $T\otimes\tilde{T}_D$ a mixture of unitary channels?

\subparagraph{Acknowledgments.} We thank R.$\F$. Werner for
valuable discussions and the $\F$, and J.I. Cirac for many useful comments and the steady support along the way. M.M.W. acknowledges financial support by QUANTOP and the Danish Natural Science Research Council (FNU).

\section{Appendix A--Matrix optimization problems}

\subsection{Minimizing $\tr[A\conj{A}]$ subject to fixed singular values}\label{sec:AAbar}

We solve the minimization problem posed in section~\ref{sec:sep_witness} which handles a.o.\ a special class of separation witnesses. Considering any complex $d \times d$ matrix $A$, we denote its singular values by $\sigma_i(A)$ (counting multiplicity) such that
$\sigma_1(A) \ge \dots \ge \sigma_d(A)$. Our main result is the following proposition, which is similar to theorem~7.4.10 in~\cite{HJ90}.
\begin{proposition}
Let $A \in \mathcal{M}_d$ and $\sigma_1 \ge \dots \ge \sigma_d$ denote the singular values of $A$. If $A \conj{A}$ is Hermitian, then there exists a permutation $\tau$ of $\left\{1, \dots, d\right\}$, an even $r \le d$ and a function $\rho: \left\{1,\dots,r/2\right\} \to \left\{0, 1\right\}$ with
\be
\tr \left[A \conj{A}\right] = 2 \sum_{i=1}^{r/2} (-1)^{\rho(i)} \, \sigma_{\tau(2i-1)} \, \sigma_{\tau(2i)} + \sum_{i=r+1}^d \sigma_{\tau(i)}^2.
\label{eq:traceAConjAFormula}
\ee
Conversely, given any such $\tau, r, \rho$ and nonnegative numbers $\sigma_1 \ge \dots \ge \sigma_d$, there exists an $A \in \mathcal{M}_d$ such that $\sigma_i(A) = \sigma_i$ for all $i$ and~\eqref{eq:traceAConjAFormula} holds.
\label{prop:traceAConjA}
\end{proposition}
\begin{proof}
We split the "$\Rightarrow$" part into the following steps:
\begin{enumerate}
\item $\tr \left[A \conj{A}\right]$ and the singular values of $A$ are invariant under $A \mapsto U A U^T$ for any unitary $U$. Note that this map sends $A \conj{A} \mapsto U A \conj{A} U^\dagger$, so by the spectral theorem, w.l.o.g. $A \conj{A}$ real diagonal.
\item It follows that $A \conj{A} = \conj{A \conj{A}} = \conj{A} A$, i.e. $A$ commutes with $\conj{A}$, and each of the eigenspaces of $A \conj{A}$ is invariant under $A$ and $\conj{A}$. Stated differently, $A$ is block diagonal, each block corresponding to an eigenspace of $A \conj{A}$, so w.l.o.g. $A\conj{A} = \lambda\,\1$ for a $\lambda \in \R$.
\item The case $\lambda \neq 0$: Applying the singular value decomposition yields unitary matrices $U, V$ such that $A_1 := U^\dagger A V = \mathrm{diag}\left(\sigma_1, \dots, \sigma_d\right)$. Using $\conj{A} = \lambda A^{-1}$, we have
\bes
A_2 := V^\dagger \conj{A} U = \lambda \left(U^\dagger A V\right)^{-1} = \lambda A_1^{-1} = \lambda \, \mathrm{diag}\left(\sigma_1^{-1}, \dots, \sigma_d^{-1}\right).
\ees
$A_1$ and $A_2$ sharing the same singular values translates to $\left\{\sigma_1,\dots,\sigma_d\right\} = \abs{\lambda} \left\{\sigma_1^{-1},\dots,\sigma_d^{-1}\right\}$, so there is a permutation $\tau$ of $\left\{1, \dots, d\right\}$ with
\bes
\left\{\begin{array}{ll}
\sigma_{\tau(2i-1)}\,\sigma_{\tau(2i)} = \abs{\lambda} \text{ for } i = 1,\dots,\frac{d}{2},& d \text{ even}\\
\sigma_{\tau(2i-1)}\,\sigma_{\tau(2i)} = \abs{\lambda} \text{ for } i = 1,\dots,\frac{d-1}{2}, \ \sigma_{\tau(d)}^2 = \abs{\lambda},& d \text{ odd}.\\
\end{array}\right.
\ees
Note that $d$ cannot be odd if $\lambda < 0$ as the negative eigenvalues of $A \conj{A}$ are of even algebraic multiplicity (see e.g.~\cite{HJ90}, pages 252, 253). Concluding, $\tr \left[A \conj{A}\right]$ can always be written in the form~\eqref{eq:traceAConjAFormula}.
\item The case $\lambda = 0$: Let $r$ denote the number of nonzero singular values of $A$. $A \conj{A} = 0$ means that $\mathrm{range}\left(\conj{A}\right) \subseteq \mathrm{kern}(A)$, so
\bes
r = \mathrm{rank}(A) = \dim \mathrm{range}\left(\conj{A}\right) \le \dim \mathrm{kern}(A) = d-r,
\ees
i.e. $2 r \le d$, and there is a permutation $\tau$ such that each summand in the right hand side of~\eqref{eq:traceAConjAFormula} is zero.
\end{enumerate}
To prove the ``$\Rightarrow$`` part, set $A := \mathrm{diag}\left(A_1, \dots, A_{r/2}, \sigma_{\tau(r+1)}, \dots, \sigma_{\tau(d)}\right)$ with
\bes
A_i := \left(\begin{smallmatrix}0&(-1)^{\rho(i)}\sigma_{\tau(2i)}\\\sigma_{\tau(2i-1)}&0\end{smallmatrix}\right).
\ees
\end{proof}

\begin{corollary}
Given any nonnegative numbers $\sigma_1 \ge \dots \ge \sigma_d$,
\be
\begin{split}
&\min \left\{\tr\left[A \conj{A}\right] \suchthat A \in \mathcal{B}\left(\mathcal{H}\right), \, \sigma_i(A) = \sigma_i \,\forall\, i \right\}\\
&= \left\{\begin{array}{ll}
-2 \sum_{i=1}^{d/2} \sigma_{2i-1} \sigma_{2i},& d \text{ even}\\
-2 \sum_{i=1}^{d-1/2} \sigma_{2i-1} \sigma_{2i} + \sigma_d^2,& d \text{ odd}.\\
\end{array}\right.\\
\end{split}
\label{eq:minTrAConjA}
\ee
\end{corollary}
\begin{proof}
What remains to be shown is $A \conj{A}$ being Hermitian for optimal $A$; then proposition~\ref{prop:traceAConjA} guarantees optimality. Exploiting invariance under $A \mapsto V A V^T$ for unitary $V$ we can w.l.o.g. assume that $A = U D$ with $D = \mathrm{diag}\left(\sigma_1, \dots, \sigma_d\right)$ and $U$ unitary. Now vary $U$ to minimize $\tr\left[A \conj{A}\right]$; the unitary constraint translates via
\bes
U\,\ud U^\dagger + \ud U\,U^\dagger = \ud \left(U\,U^\dagger\right) = 0
\ees
to $X := i \ud U\,U^\dagger$ being Hermitian. We have
\bes
\begin{split}
0 &\stackrel{!}{=}\,\frac{\ud}{\ud U} \tr\left[D U D \conj{U}\right]\\
&= 2\,\Re\,\tr\left[D \ud U D \conj{U}\right]\\
&= 2\,\Im\,\tr\left[U D \conj{U} D X\right].
\end{split}
\ees
This has to hold true for any Hermitian matrix $X$. Decomposing $U D \conj{U} D = B_1 + i B_2$, $B_1$ and $B_2$ Hermitian, it follows that $B_2 = 0$.
\end{proof}

\subsection{Maximizing $|\tr[U]|$ subject to fixed $\tr[U\conj{U}]$}\label{sec:maxTrUnitary}

In this subsection we prove proposition~\ref{prop:traceUvsTraceUconjU1} from section~\ref{sec:completepic}, i.e.\ we calculate the analytic solution of $\max\abs{\tr\,U}$ for fixed $\tr\left[U\conj{U}\right]$ over all unitaries $U \in U(d)$. The motivation for this optimization problem comes from Eq.~\eqref{eq:eval_orthcov_unitary}, which characterizes the convex hull of unitary channels within the set of orthogonal-covariant channels.

We need the following lemma first.
\begin{lemma}
Let $U \in U(2)$ be a unitary $2 \times 2$ matrix and $U_\mathfrak{s} := \frac{1}{2}\left(U + U^T\right)$ the \emph{symmetric} part of $U$. Then the trace-norm of $U_\mathfrak{s}$ equals
\bes
\norm{U_\mathfrak{s}}_1 \equiv \sum_{j=1,2} \sigma_j\left(U_\mathfrak{s}\right) = \sqrt{\tr\left[U \conj{U}\right]+2}.
\ees
\label{lem:Dim2TraceUvsTraceUU}
\end{lemma}
\begin{proof}
There are $\alpha, \beta \in \C$ such that up to an unimportant phase factor
\bes
U = \begin{pmatrix}\alpha&\beta\\-\conj{\beta}&\conj{\alpha}\end{pmatrix} \quad \text{with} \quad \abs{\alpha}^2 + \abs{\beta}^2 = 1.
\ees
Direct calculation shows that
\begin{align}
\tr\left[U \conj{U}\right] &= 2 \left(\abs{\alpha}^2 - \Re\left(\beta^2\right)\right) = 4 \left(\abs{\alpha}^2 + (\Im\,\beta)^2\right) - 2,
\label{eq:Dim2TraceUConjU}\\
U_\mathfrak{s} &= \begin{pmatrix}\alpha&i\,\Im\,\beta\\i\,\Im\,\beta&\conj{\alpha}\end{pmatrix}, \quad U_\mathfrak{s} U_\mathfrak{s}^\dagger = \left(\abs{\alpha}^2 + (\Im\,\beta)^2\right) \1,\label{eq:Dim2SigmaSU}\\
\sum_{j=1,2} \sigma_j\left(U_\mathfrak{s}\right) &\stackrel{\eqref{eq:Dim2SigmaSU}}{=} 2 \sqrt{\abs{\alpha}^2 + (\Im\,\beta)^2} \stackrel{\eqref{eq:Dim2TraceUConjU}}{=} \sqrt{\tr\left[U \conj{U}\right]+2}.\notag
\end{align}
\end{proof}

\begin{proposition}
Let $d \ge 1$ be odd. Then for all $x \in \left[-1+\frac{2}{d}, 1\right]$, there exists a unitary $U \in U(d)$ such that $d^{-1} \tr\left[U \conj{U}\right] = x$, and
\be
\begin{split}
&\max\left\{d^{-1} \abs{\tr\,U} \suchthat U \in U(d),\,d^{-1} \tr\left[U \conj{U}\right] = x\right\}\\
&= \left[\frac{1}{2} \left(1 - \frac{1}{d}\right) \left(1 - \frac{2}{d} + x\right)\right]^{1/2} + \frac{1}{d}.\\
\end{split}
\label{eq:TraceUvsTraceUConjU}
\ee
\label{prop:TraceUvsTraceUConjU}
\end{proposition}
\begin{proof}
Set $\alpha := \frac{d}{d-1} \left[\frac{1}{2} \left(1-\frac{1}{d}\right) \left(1-\frac{2}{d}+x\right)\right]^{1/2} \in [0,1]$, $\beta := \sqrt{1-\alpha^2}$ and $\sigma := \left(\begin{smallmatrix}\alpha&\beta\\-\beta&\alpha\end{smallmatrix}\right)$, then a short calculation shows that $U := \mathrm{diag} \left(\sigma, \dots, \sigma, 1\right)$ satisfies $d^{-1} \tr\left[U \conj{U}\right] = x$ with $d^{-1} \abs{\tr\,U}$ equal to the right hand side of~\eqref{eq:TraceUvsTraceUConjU}. So what remains to be shown is an upper bound on $d^{-1} \abs{\tr\,U}$.

Let $U \in U(d)$ be a unitary matrix with $d^{-1} \tr\left[U \conj{U}\right] = x$. By the Youla theorem~\cite{FI06}, given any \emph{conjugate-normal} matrix $A$ (that is, $A A^\dagger = \conj{A^\dagger A}$), there exists a unitary $V$ such that $V A V^T$ is a block diagonal matrix with diagonal blocks of order $1 \times 1$ and $2 \times 2$, the $1 \times 1$ blocks corresponding to the real nonnegative eigenvalues of $A \conj{A}$ and the $2 \times 2$ blocks corresponding either to pairs of equal negative eigenvalues of $A \conj{A}$ or to conjugate pairs of non-real eigenvalues of $A \conj{A}$. Applying this to $U$, there is a unitary $V$ with $U = V D V^T$, the block diagonal matrix $D$ as described. Let $\frac{r}{2}$ be the number of $2 \times 2$ blocks (with even $r \le d$) and denote these blocks by $D_i$. As $U$ and $V$ are unitary, so must be $D$, i.e. $D_i \in U(2)$ for all $i$. $U \conj{U}$ unitary guarantees $\abs{\lambda} = 1$ for each eigenvalue $\lambda$ of $U \conj{U}$, so each real nonnegative eigenvalue of $U \conj{U}$ must be $1$. Altogether we have $D = \mathrm{diag}\left(D_1, \dots, D_{\frac{r}{2}}, 1, \dots, 1\right)$. Set $c_i := \frac{1}{2}\,\tr\left[D_i \conj{D_i}\right] \in [-1,1]$ and $D_\mathfrak{s} := \frac{1}{2}\left(D + D^T\right)$. Using Lemma~\ref{lem:Dim2TraceUvsTraceUU} and the fact that $V^T V$ is unitary and symmetric,
\bes
\begin{split}
\abs{\tr\,U}
&= \abs{\tr\left[D V^T V\right]} = \abs{\tr\left[D_\mathfrak{s} V^T V\right]} \le \sum_{j=1}^d \sigma_j\left(D_\mathfrak{s}\right)\\
&= \sum_{i=1}^{r/2} \sum_{j=1,2} \sigma_j\left(D_{i,\mathfrak{s}}\right) + (d - r) = 2 \cdot \sum_{i=1}^{r/2} \sqrt{\frac{c_i+1}{2}} + d - r.
\end{split}
\ees
Let $c := d \cdot x \in [-d + 2, d]$. Some elementary analysis shows that the problem
\begin{align*}
\max&\left\{2 \cdot \sum_{i=1}^{r/2} \sqrt{\frac{c_i+1}{2}} + d - r\right\}\\
\text{subject to}\quad
&2 \cdot \sum_{i=1}^{r/2} c_i + d - r = c,\\
&c_i \in [-1,1] \quad \text{for all } i,\\
&r \text{ even},\, r \le d
\end{align*}
has optimal solution $r = d-1$, $c_i = \frac{c-1}{d-1} \,\forall\,i$ and the obtained maximum is $d \left[\frac{1}{2} \left(1-\frac{1}{d}\right) \left(1-\frac{2}{d}+\frac{c}{d}\right)\right]^{1/2} + 1$. This upper bound on $\abs{\tr\,U}$ corresponds exactly to the right hand side of~\eqref{eq:TraceUvsTraceUConjU}.
\end{proof}

\subsection{Minimizing $\tr\big[U_\mathfrak{s}\,\conj{U_\mathfrak{s}}^{T_2}\big]$ subject to fixed $\tr[U\conj{U}]$}\label{sec:minTrPTSymU}

Motivated by Eq.~\eqref{eq:uig} in section~\ref{sec:2copies}, we investigate
\be
\begin{split}
&\min\left\{\frac{1}{d_1 d_2}\tr\left[U_\mathfrak{s}\,\conj{U_\mathfrak{s}}^{T_2}\right] \suchthat U_\mathfrak{s} = \frac{1}{2}\left(U + U^T\right),\right.\\
&\qquad\quad\left.U \in \mathcal{B}\left(\mathcal{H}\otimes\mathcal{K}\right) \text{ unitary with } \frac{1}{d_1 d_2}\tr\left[U \conj{U}\right] = y\right\}
\label{eq:minTrPTSymU}
\end{split}
\ee
for Hilbert spaces $\mathcal{H}$ and $\mathcal{K}$ with dimensions $d_1$ and $d_2$, respectively, and provided $y \in \left[-1+2/d_1 d_2,1\right]$, where $d_1 \ge 3$ is odd.

The partial transposes $^{T_1}$ and $^{T_2}$ are defined w.r.t. a fixed product basis by the linear extension of
\bes
\begin{split}
\left(A \otimes B\right)^{T_1} &= A^T \otimes B \quad \text{and}\\
\left(A \otimes B\right)^{T_2} &= A \otimes B^T, \quad A \otimes B \in \mathcal{B}\left(\mathcal{H}\otimes\mathcal{K}\right),
\end{split}
\ees
respectively. Note that for any $A$ and $B$,
\bes
\tr\left[A\,\conj{B}^{T_2}\right] = \tr\left[A^{T_1}\,B^\dagger\right],
\ees
and for any real or complex $A$ with $A^T = A$, the partial transposes are on equal footing, i.e. $A^{T_1} = A^{T_2}$, so~\eqref{eq:minTrPTSymU} is  inherently \emph{symmetric} with respect to $d_1 \leftrightarrow d_2$. We identify $\mathcal{B}\left(\mathcal{H}\otimes\mathcal{K}\right) \cong \C^{d_1 d_2 \times d_1 d_2}$ by means of the ordered computational basis $\left(\ket{11}, \ket{12},\dots\ket{1 d_2},\dots\ket{d_1 d_2}\right)$.

All quantities in~\eqref{eq:minTrPTSymU}, especially the minimizers, will stay invariant if we send
\bes
U \to \left(W_1 \otimes W_2\right) U \left(W_1 \otimes W_2\right)^T
\ees
with arbitrary unitaries $W_1 \in \mathcal{B}(\mathcal{H})$ and $W_2 \in \mathcal{B}(\mathcal{K})$.

Since every unitary matrix is also conjugate-normal (that is, $U\,U^\dagger = \conj{U^\dagger\,U}$), the Youla-theorem\footnote{Refer in particular to Thm.~4 in~\cite{FI06}. The Youla-form corresponds to the Schur-form w.r.t. unitary congruence transformations $A \mapsto V A V^T$ and is a generalization of Takagi's factorization~\cite{HJ90}.} states that there exists a unitary matrix $V \in \mathcal{B}\left(\mathcal{H}\otimes\mathcal{K}\right)$ such that
\bes
U = V\,D\,V^T
\ees
with $D$ \emph{real} block-diagonal and blocks of size $1 \times 1$ and $2 \times 2$, the former non-negative and the latter of the form $\left(\begin{smallmatrix}\sigma&-z\\z&\sigma\end{smallmatrix}\right)$ with $\sigma \ge 0$. Since $D$ must also be unitary, this equals $\left(\begin{smallmatrix}\cos\vartheta&-\sin\vartheta\\\sin\vartheta&\cos\vartheta\end{smallmatrix}\right)$ for a $\vartheta \in \R$, and all $1 \times 1$ blocks are $1$. Note that
\bes
U_\mathfrak{s} = V\,D_\mathfrak{s}\,V^T, \quad D_\mathfrak{s} := \frac{1}{2}\left(D + D^T\right) \ge 0\, \text{ diagonal},
\ees
in particular, $D_\mathfrak{s}$ contains the singular values of $U_\mathfrak{s}$. Moreover, $\tr\left[U\conj{U}\right] \equiv \tr\left[D\conj{D}\right]$ is independent of $V$, so $D$ fixes $y$ in~\eqref{eq:minTrPTSymU} and we may freely vary $V$. Conversely, Takagi's theorem~\cite{HJ90} asserts that every complex-symmetric matrix $A \in \C^{n \times n}$ can be decomposed into
\be A = V\,\mathrm{diag}\left(\sigma_1,\dots,\sigma_n\right)\,V^T \label{eq:TakagiA}\ee
with unitary $V$ and $\sigma_i \ge 0$ for all $i$, so identifying $A \equiv U_\mathfrak{s}$ and $\mathrm{diag}\left(\sigma_1,\dots,\sigma_n\right) \equiv D_\mathfrak{s}$, the minimization problem~\eqref{eq:minTrPTSymU} can be reduced to the following problem and a subsequent optimization over $D_\mathfrak{s}$:
\be
\min\left\{\frac{1}{d_1 d_2} \tr\left[A\,\conj{A}^{T_2}\right] \suchthat A \in \mathcal{B}(\mathcal{H}\otimes\mathcal{K}),\, A^T = A,\, \sigma_i(A) = \sigma_i\,\forall\,i\right\}.
\label{eq:minTrPTSymA}
\ee
This closely resembles~\eqref{eq:minTrAConjA}, and we have effectively decoupled the target function from the peculiar unitary constraint in~\eqref{eq:minTrPTSymU}.

\begin{proposition}
Every (local) minimizer $A$ of~\eqref{eq:minTrPTSymA} satisfies $A\,\conj{A}^{T_2}$ Hermitian.
\end{proposition}
\begin{proof}
Denote the derivative w.r.t. $V$ in~\eqref{eq:TakagiA} by $\ud V$; since $V$ is unitary,
\bes
V\,\ud V^\dagger + \ud V\,V^\dagger = \ud \left(V\,V^\dagger\right) = 0
\ees
so $X := \frac{1}{i}\ud V\,V^\dagger$ must be Hermitian. Plugging
\bes
\frac{\ud}{\ud V} A = \ud V\,\mathrm{diag}\left(\sigma_i\right)\,V^T + V\,\mathrm{diag}\left(\sigma_i\right)\,\ud V^T = i \left(X A + A X^T\right)
\ees
into the target function~\eqref{eq:minTrPTSymA} yields
\bes
\begin{split}
\frac{\ud}{\ud V} \tr\left[A\,\conj{A}^{T_2}\right]
&= i\,\tr\left[\left(X A + A X^T\right)\,\conj{A}^{T_2}\right]\\
&- i\,\tr\left[A\,\conj{\left(X A + A X^T\right)}^{T_2}\right]\\
&= 2\,i\,\tr\left[X\,A\,\conj{A}^{T_2}\right]
- i\,\tr\left[A^{T_2}\,\left(X A + A X^T\right)^\dagger\right]\\
&= 2\,i\,\tr\left[X \left(A\,\conj{A}^{T_2} - A^{T_2}\,\conj{A}\right)\right]\\
&= 2\,i\,\tr\left[X \left(A\,\conj{A}^{T_2} - \text{h.c.}\right)\right] \stackrel{!}{=} 0.\\
\end{split}
\ees
As this must hold for any Hermitian $X$, the last equation can only be fulfilled if $A\,\conj{A}^{T_2}$ is Hermitian.
\end{proof}

It is instructive to rewrite the target function as follows, setting $\sigma := \left(\sigma_1,\dots,\sigma_{d_1 d_2}\right)$. Denote the columns of $V$ by $v_i$, i.e. $V = \left(v_1 \vert\dots\vert v_{d_1 d_2}\right)$, then $A = \sum_i \sigma_i\,v_i v_i^T$ and
\be\begin{split}
\frac{1}{d_1 d_2}\tr&\left[A\,\conj{A}^{T_2}\right] = \hprod{\sigma}{G\,\sigma},\\
G &= \left(g_{ij}\right) \quad \text{Hermitian with}\\
g_{ij} &= \frac{1}{d_1 d_2} \tr\left[v_j v_j^T\,\conj{v_i v_i^T}^{T_2}\right].
\end{split}
\label{eq:splitDandVbyG}\ee
Writing
$
v_i =: \sum_k \ket{k}\otimes x_{ik}, \quad x_{ik} \in \mathcal{K}
$
the last expression becomes
\bes
\begin{split}
g_{ij}
&= \frac{1}{d_1 d_2} \sum_{k,k',l,l'} \tr\left[\left(\ket{l'}\bra{l}\otimes x_{jl'} x_{jl}^T\right) \left(\ket{k'}\bra{k}\otimes \left(x_{ik'} x_{ik}^T\right)^\dagger\right)\right]\\
&= \frac{1}{d_1 d_2} \sum_{k,l} \tr_\mathcal{K}\left[x_{jk}\,x_{jl}^T\,\conj{x_{ik}}\,x_{il}^\dagger\right]\\
&= \frac{1}{d_1 d_2} \sum_{k,l} \hprod{x_{ik}}{x_{jl}} \hprod{x_{il}}{x_{jk}}\\
&= \frac{1}{d_1 d_2} \tr\left[s_{ij}^2\right], \quad s_{ij} := \left(\hprod{x_{ik}}{x_{jl}}\right)_{k,l=1,\dots, d_1}.
\end{split}
\ees
$V$ being unitary translates to
$
\tr\,s_{ij} = \sum_k \hprod{x_{ik}}{x_{jk}} = \hprod{v_i}{v_j} \stackrel{!}{=} \delta_{ij}.
$

In what follows, we provide an explicit upper bound\footnote{Numeric tests strongly suggest that this is the actual minimum. Most interestingly, the acclaimed minimizer $V$ does not depend on $\vartheta$!} of~\eqref{eq:minTrPTSymU} for $d_1 = d_2 =: d$ and $d = 3$. Start with the \emph{Ansatz} that all $2 \times 2$ blocks in $D$ belong to the same phase, i.e.
\bes
D = \begin{pmatrix}\cos\vartheta&-\sin\vartheta\\\sin\vartheta&\cos\vartheta\end{pmatrix}^{\otimes 4} \oplus 1, \quad \vartheta \in \left[0,\pi/2\right], \quad \text{so}
\ees
\be
\frac{1}{d^2}\tr\left[U\conj{U}\right] = \frac{1}{d^2}\tr\left[D\conj{D}\right] = \frac{1}{9}\left(16\cos^2\vartheta-7\right) \stackrel{!}{=} y,
\label{eq:ThetaY}\ee
and set
\bes
V = \left(\begin{array}{ccc|ccc|ccc}
&\frac{i}{\sqrt{2}}&-\frac{i}{2\sqrt{6}}&&-\frac{i}{2 \sqrt{2}}&&&&\frac{1}{\sqrt{3}}\\
\frac{i}{\sqrt{2}}&&-\sqrt{\frac{3}{8}}&&\frac{1}{2 \sqrt{2}}&&&&\\
&&&&&\frac{1}{\sqrt{2}}&&\frac{i}{\sqrt{2}}&\\
\hline
\frac{i}{\sqrt{2}}&&\sqrt{\frac{3}{8}}&&-\frac{1}{2 \sqrt{2}}&&&&\\
&&\frac{i}{\sqrt{6}}&&\frac{i}{\sqrt{2}}&&&&\frac{1}{\sqrt{3}}\\
&&&\frac{1}{\sqrt{2}}&&&\frac{i}{\sqrt{2}}&&\\
\hline
&&&&&-\frac{1}{\sqrt{2}}&&\frac{i}{\sqrt{2}}&\\
&&&-\frac{1}{\sqrt{2}}&&&\frac{i}{\sqrt{2}}&&\\
&-\frac{i}{\sqrt{2}}&-\frac{i}{2\sqrt{6}}&&-\frac{i}{2\sqrt{2}}&&&&\frac{1}{\sqrt{3}}\\
\end{array}\right)
\ees
independent of $\vartheta$! Then $D_\mathfrak{s} = \mathrm{diag}\left(\sigma_1,\dots,\sigma_9\right)$ with $\sigma_1 = \dots = \sigma_8 = \cos\vartheta$, $\sigma_9 = 1$, and $G$ in~\eqref{eq:splitDandVbyG} becomes
\bes
G = \frac{1}{27}\left(\begin{array}{ccc|ccc|ccc}
\frac{3}{2}&0&-{\frac{11}{8}}&0&-{\frac{9}{8}}&0&0&0&-1\\
0&\frac{3}{2}&\frac{1}{8}&0&\frac{3}{8}&-\frac{3}{2}&0&-\frac{3}{2}&-1\vphantom{\Bigl(}\\
-\frac{11}{8}&\frac{1}{8}&\frac{3}{2}&-\frac{1}{4}&-1&\frac{1}{8}&-\frac{1}{4}&\frac{1}{8}&-1\vphantom{\Bigl(}\\
\hline
0&0&-\frac{1}{4}&\frac{3}{2}&-\frac{3}{4}&0&-\frac{3}{2}&0&-1\vphantom{\Bigl(}\\
-\frac{9}{8}&\frac{3}{8}&-1&-\frac{3}{4}&\frac{3}{2}&\frac{3}{8}&-\frac{3}{4}&\frac{3}{8}&-1\\
0&-\frac{3}{2}&\frac{1}{8}&0&\frac{3}{8}&\frac{3}{2}&0&-\frac{3}{2}&-1\vphantom{\Bigl(}\\
\hline
0&0&-\frac{1}{4}&-\frac{3}{2}&-\frac{3}{4}&0&\frac{3}{2}&0&-1\vphantom{\Bigl(}\\
0&-\frac{3}{2}&\frac{1}{8}&0&\frac{3}{8}&-\frac{3}{2}&0&\frac{3}{2}&-1\vphantom{\Bigl(}\\
-1&-1&-1&-1&-1&-1&-1&-1&1\\
\end{array}\right) \in \Q^{9 \times 9}.
\ees
Finally evaluating the target function provides the supposed minimum
\be
\frac{1}{d^2}\tr\left[U_\mathfrak{s}\,\conj{U_\mathfrak{s}}^{T_2}\right] = \hprod{\sigma}{G\,\sigma} = \frac{1}{9}\left(-\frac{8}{3}\left(\cos\vartheta+1\right)^2+3\right)
\label{eq:minTrPTSymUTheta}\ee
with $\vartheta$ defined by~\eqref{eq:ThetaY}.

Interestingly, the smallest eigenvalue $-\frac{1}{d^2}$ of $G$ is of algebraic multiplicity $1$ with corresponding eigenvector $\sigma(\vartheta)$ evaluated at $\vartheta = \pi/3$ and coordinates $-\frac{1}{3}\left(1,1\right)$. Furthermore, \eqref{eq:minTrPTSymUTheta}~is minimal w.r.t. $\vartheta$ exactly for $\vartheta = 0$, which corresponds to maximal $y = \frac{1}{d^2}\tr\left[U \conj{U}\right] = 1$ and $\sigma_1 = \dots = \sigma_{d^2} = 1$. In this case, $D$ is the identity and $U = V\,V^T$ complex symmetric. Comparing with appendix~\ref{sec:minTrPTUnitary}, notice that we obtain the same minimum value $-\frac{23}{27}$.

\subsection{Minimizing $\tr\big[U\,\conj{U}^{T_2}\big]$}\label{sec:minTrPTUnitary}

We explore the following minimization problem posed by Eq.~\eqref{CoordDepU} in section~\ref{sec:noisy}.
\be
\min \left\{\frac{1}{d_1 d_2} \tr\left[U\,\conj{U}^{T_2}\right] \suchthat U \in \mathcal{B}\left(\mathcal{H}\otimes\mathcal{K}\right) \text{ unitary}\right\}
\label{eq:minTrPTUnitary}
\ee
where $\mathcal{H}\otimes\mathcal{K}$ is the tensor product of two Hilbert spaces with dimensions $d_1 = \dim \mathcal{H}$ and $d_2 = \dim \mathcal{K}$, respectively, $d_1$ being odd. The partial transposition is introduced in~\ref{sec:minTrPTSymU}. Note that any transformation
\bes U \to \left(V^T \otimes W_1^\dagger\right) U \left(V \otimes W_2^{\vphantom{\dagger}}\right)
\ees
for unitary $V \in \mathcal{B}\left(\mathcal{H}\right)$ and unitary $W_1, W_2 \in \mathcal{B}\left(\mathcal{K}\right)$ leaves the target function invariant. If we allowed tensor products only, i.e. $U = U_1 \otimes U_2$, the target function would collapse to $\frac{1}{d_1}\,\tr\left[U_1 \conj{U_1}\right] \ge -1 + \frac{2}{d_1}$, which is in general strictly greater than~\eqref{eq:minTrPTUnitary}, see below. It is worth mentioning that~\eqref{eq:minTrPTUnitary} is inherently asymmetric w.r.t. $d_1 \leftrightarrow d_2$, as opposed to the previous section~\ref{sec:minTrPTSymU}.

\begin{proposition}
$U\,\conj{U}^{T_2}$ is Hermitian for every minimizer $U$ of~\eqref{eq:minTrPTUnitary}.
\label{prop:hermitianMinTrPTU}
\end{proposition}
\begin{proof}
As in previous sections, we differentiate the target function with respect to $U$. As $U$ is unitary, $X := \frac{1}{i}\ud U\,U^\dagger$ must be Hermitian, and we get
\bes
\begin{split}
\frac{\ud}{\ud U} \tr\left[U\,\conj{U}^{T_2}\right]
&= i\,\tr\left[X U \conj{U}^{T_2}\right] - i\,\tr\left[U^{T_1} \left(X U\right)^\dagger\right]\\
&= i\,\tr\left[X \left(U \conj{U}^{T_2} - \text{h.c.}\right)\right] \stackrel{!}{=} 0.
\end{split}
\ees
This holds for any Hermitian $X$, so $U \conj{U}^{T_2}$ must be Hermitian, too.
\end{proof}

\subparagraph{Disassembly and reformulation.}

Let $\mathcal{X} = \mathcal{B}\left(\mathcal{K}\right)$ be the Hilbert space equipped with the Hilbert-Schmidt inner product
and induced Frobenius norm.
By partitioning $U$ as $U = \sum_{i,j=1}^{d_1} \ket{i}\bra{j} \otimes u_{ij}$ with $u_{ij} \in \mathcal{X}$, we can now reformulate the target function~\eqref{eq:minTrPTUnitary} as
\bes
\tr\left[U\,\conj{U}^{T_2}\right] = \sum_{i,j} \hprod{u_{ij}}{u_{ji}}.
\ees
The condition $U$ unitary translates to
\bes
\begin{split}
U U^\dagger &= \1 \quad \Leftrightarrow \quad \sum_i u_{ij}^\dagger u_{ik} = \delta_{jk}\1 \quad \forall\,j,k = 1,\dots,d_1\\
&\Leftrightarrow \quad \sum_i \hprod{u_{ij}}{u_{ik}\,x} = \delta_{jk}\hprod{\1}{x} \equiv \delta_{jk}\,\tr\,x \quad \forall\,j,k;\, x \in \mathcal{X}
\end{split}
\ees
and the condition in proposition~\ref{prop:hermitianMinTrPTU} to
\bes
\begin{split}
U\,\conj{U}^{T_2}& \text{ Hermitian} \quad \Leftrightarrow \quad \sum_i u_{ki}\,u_{ij}^\dagger = \sum_i u_{ik}\,u_{ji}^\dagger\\
&\Leftrightarrow \quad \sum_i \hprod{u_{ij}}{x\,u_{ki}} = \sum_i \hprod{u_{ji}}{x\,u_{ik}} \quad \forall\,j,k;\, x \in \mathcal{X}.
\end{split}
\ees
Note that these equations can be rewritten in terms of the Hilbert-Schmidt inner product as shown.

\subparagraph{Quaternion structure.}

In this paragraph we assume $d_2 = 2$ and set $d = d_1$. Numeric tests suggest that in this case, the minimum value~\eqref{eq:minTrPTUnitary} is
\be
-1+\frac{2}{d^2}.
\label{eq:minNumTrPTD2}
\ee
Interestingly, there emerges a substructure which is best described by quaternions. Recall that quaternions
\bes
\quat = \left\{x_0 + x_1\,i + x_2\,j + x_3\,k \suchthat x_0, \dots, x_3 \in \R\right\}
\ees
are a non-abelian division ring and form a 4-dimensional normed division algebra over the real numbers. We regard $\R$ and $\C$ as subalgebras of $\quat$ and denote the quaternion-conjugate of $q = x_0 + x_1\,i + x_2\,j + x_3\,k \in \quat$ by $q^*$. Furthermore, define $\Re\,q := x_0$ and $\vec{q} := q - \Re\,q = x_1\,i + x_2\,j + x_3\,k$.

To bridge the gap between quaternions and operators on Hilbert spaces, employ the identification\footnote{We adhere here to a different convention than e.g.~\cite{GerstnerByersMehrmann1989}.}
\bes
i \leftrightarrow i\,\sigma_z = \begin{pmatrix}i&0\\0&-i\end{pmatrix}, \quad
j \leftrightarrow i\,\sigma_y = \begin{pmatrix}0&1\\-1&0\end{pmatrix}, \quad
k \leftrightarrow i\,\sigma_x = \begin{pmatrix}0&i\\i&0\end{pmatrix}.
\ees
Note that in this representation
\be
q \leftrightarrow \hat{q} = \begin{pmatrix}x_0 + i\,x_1&x_2 + i\,x_3\\-x_2 + i\,x_3&x_0 - i\,x_1\end{pmatrix} \in \C^{2 \times 2}
\label{eq:quaternionSU2}
\ee
the quaternion conjugate is the Hermitian conjugate of the corresponds matrix, and the quaternion norm is the square root of the determinant,
\bes q^* \leftrightarrow \hat{q}^\dagger, \quad \norm{q} = \sqrt{\det(\hat{q})}.\ees

Let $\quat^d$ be the $d$-dimensional ``vector space'' over $\quat$ with multiplication from the right, then each linear transformation on $\quat^d$ can be represented by a $d \times d$ matrix $A \in \quat^{d \times d}$. The identification~\eqref{eq:quaternionSU2} provides an algebra isomorphism between $\quat^{d \times d} \cong \R^{d \times d} \otimes \quat$ and the complex $2d \times 2d$ matrices consisting of $2 \times 2$ blocks~\eqref{eq:quaternionSU2}; to obtain $A x$, define $u, v \in \C^d$ by $x =: u - j\,v$ and set $\hat{x} = \left(u_1,v_1,\dots,u_d,v_d\right)^T \in \C^{2d}$; then $A x$ corresponds exactly to $\hat{A}\hat{x}$. In the following it will be clear from context which representation is employed.

For all $A \in \quat^{d \times d}$, the component-wise quaternion conjugate $A^*$ and the quaternion conjugate-transpose $A^\dagger$ are intuitively translated to
\bes
A^* \leftrightarrow \conj{\hat{A}}^{T_2}, \quad A^\dagger \leftrightarrow \hat{A}^\dagger.
\ees
Consequently, we say that $A$ is Hermitian if $A^\dagger = A$.

As in~\cite{GerstnerByersMehrmann1989}, call $\lambda \in \C$, $\Im\,\lambda \ge 0$ an eigenvalue of $A$ with corresponding eigenvector $x \in \quat^d$ if $A x = x \lambda$. These are exactly the eigenvalues of $\hat{A}$ which have nonnegative imaginary part. Note that most of the well-known linear algebra theorems can be generalized straightforward to quaternions.

\begin{proposition}
Let $A \in \quat^{d \times d}$ be Hermitian with eigenvalues $-1 \pm d$, the algebraic multiplicity of $(-1+d)$ being $\frac{d+1}{2}$ and of $-(1+d)$ being $\frac{d-1}{2}$, respectively. Suppose $A$ can be chosen such that $\Re\,A = 0$, then there exists a unitary $U \in \mathcal{B}\left(\mathcal{H}\otimes\mathcal{K}\right)$ with
\bes
\frac{1}{2\,d}\tr\left[U\,\conj{U}^{T_2}\right] = -1+\frac{2}{d^2}.
\ees
\label{prop:quaternionMinTrPTU}
\end{proposition}
\begin{proof}
Since $\Re\,A = 0$, we have $A^* = -A$, and consequently $A^T = -A$. Set $U = \frac{1}{d}(\1 + A)$, embedding $\quat^{d \times d}$ into $\C^{d \times d}\otimes\C^{2 \times 2}$ as described above, then $U$ will be Hermitian and unitary since the eigenvalues satisfy $\lambda(U) = \frac{1}{d}\left(1 + (-1 \pm d)\right) = \pm 1$. Using $\conj{U}^{T_2} \equiv \frac{1}{d}\left(\1 + A^*\right) = \frac{1}{d}\left(\1 - A\right)$, we get
\bes
\frac{1}{2\,d}\tr\left[U\,\conj{U}^{T_2}\right] = \frac{1}{2\,d}\frac{1}{d^2}\tr\left[\1 - A^2\right] = \frac{1}{d^2}\left(1 - \left(d^2-1\right)\right) = -1 + \frac{2}{d^2}.
\ees
The isomorphism~\eqref{eq:quaternionSU2} introduces an additional factor $2$ into the trace, which cancels $\frac{1}{2}$ in the last equation.
\end{proof}
Note that the conditions on $A$ can be rephrased as follows. $A \in \quat^{d \times d}$ is Hermitian with $\Re\,A = 0$ such that $A^2 + 2\,A - (d^2-1)\1 = 0$. The requirement $\Re\,A = 0$ implies the respective eigenvalue multiplicities via $\tr\,A = 0$.

To ensoul proposition~\ref{prop:quaternionMinTrPTU}, we provide explicit examples of $A$ meeting all requirements for $d = 3$ and $d = 5$, namely
\bes
A = 2 \begin{pmatrix}
0&-i&j\\
i&0&-k\\
-j&k&0\\
\end{pmatrix}
\ees
and
\bes
A = \begin{pmatrix}
0&-2 i&-\sqrt{12} j&2 k&-2 j\\
2 i&0&0&-2 j&4 k\\
\sqrt{12} j&0&0&-\sqrt{12} i&0\\
-2 k&2 j&\sqrt{12} i&0&-2 i\\
2 j&-4 k&0&2 i&0\\
\end{pmatrix}
\ees
respectively. Note that for these quaternion models, $U$ is Hermitian, as well as $U\,\conj{U}^{T_2}$, in agreement with proposition~\ref{prop:hermitianMinTrPTU}.

\subparagraph{Higher dimensions.}

\begin{table}[!th]
\centering
\begin{tabular}{r||rcccc}
$d_1$& $d_2 = 1$& 2& 3& 4& 5\\
\hline
1& 1& 1& 1& 1& 1 $\vphantom{\Bigl(}$\\
3& $-\frac{1}{3}$& $-\frac{7}{9}$& $-\frac{23}{27}$& $-\frac{4}{5}$& $-\frac{37}{45}$ $\vphantom{\Bigl(}$\\
5& $-\frac{3}{5}$& $-\frac{23}{25}$& $-0.92915\dots$ & ${-1}$& $-0.97632\dots$ $\vphantom{\Bigl(}$\\
7& $-\frac{5}{7}$& $-\frac{47}{49}$& & & $\vphantom{\Bigl(}$\\
\end{tabular}
\caption{Numeric solutions of~\eqref{eq:minTrPTUnitary}; the columns correspond to different values of $d_2$. The case $d_2 = 1$ is treated analytically and is a special case of Sec.~\ref{sec:AAbar}. Note that $d_2 = 2$ is in agreement with~\eqref{eq:minNumTrPTD2}, and for $d_1 = d_2 = 3$ we obtain the same value as the minimum of~\eqref{eq:minTrPTSymUTheta} with respect to $\vartheta$. }
\label{tab:minTrPTUnitary}
\end{table}

Table~\ref{tab:minTrPTUnitary} contains numerical results for different values of $d_1$ and $d_2$. We have simply employed $U = \exp\left[i\,X\right]$ with Hermitian $X$ to represent unitary matrices. The local convergence error is about $10^{-6}$, but it is still difficult to find the \emph{global} minimizers. Quite remarkably, it seems that the lower bound $-1$ can be obtained for $d_1 = 5$ and $d_2 = 4$, even if we restrict to real orthogonal matrices.

\section{Appendix B--A special extremal channel}\label{sec:extremeUnitalAppendix}

The following algebraic values for the coefficients $\mu_1, \dots, \mu_4$ of $X$ in~\eqref{eq:extremeX} are appropriate; we have obtained them basically by guessing and suppose that at least polynomial degree $3$ is required.
\bes
\begin{split}
\mu_1
&= \frac{1}{6}\,\mathrm{Root}_1\left[-356 + 312 x - 66 x^2 + 3 x^3\right]^{1/2}\\
&= \frac{1}{6}\,\left(\frac{22}{3} - \frac{43 \cdot 2^{2/3}}{3 \left(977 + 213\,i\,\sqrt{7}\right)^{1/3}} \left(1 + i\,\sqrt{3}\right) - \frac{\left(977 + 213\,i\,\sqrt{7}\right)^{1/3}}{3 \cdot 2^{2/3}} \left(1 - i\,\sqrt{3}\right)\right)^{1/2}\\
&\doteq 0.21821,\\
\mu_2
&= \mathrm{Root}_1\left[-1 + 432 x^2 + 2592 x^3\right]\\
&= -\frac{1}{18} - \frac{1 + i\,\sqrt{3}}{18 \left(1 + 3\,i\,\sqrt{7}\right)^{1/3}} - \frac{1}{72} \left(1 - i\,\sqrt{3}\right) \left(1 + 3\,i\,\sqrt{7}\right)^{1/3}\\
&\doteq -0.14937,\\
\mu_3 &= \frac{1}{6},\\
\mu_4
&= \mathrm{Root}_2\left[1 - 6 x + 18 x^3\right]\\
&= -\frac{\left(1 + i\,\sqrt{3}\right) \left(-3 + i\,\sqrt{7}\right)^{1/3}}{6 \cdot 2^{2/3}} - \frac{1 - i\,\sqrt{3}}{3 \left(2 \left(-3 + i\,\sqrt{7}\right)\right)^{1/3}}\\
&\doteq 0.18595.\\
\end{split}
\ees
The eigenvalues of $X$ as calculated by a computer algebra program are then
\bes
\begin{split}
\lambda(X)
&= \left\{0,0,\frac{1}{3},\frac{1}{3}, \frac{1 \pm \sqrt{\alpha}}{6}\right\}\\
&\doteq \left\{0,0,\frac{1}{3},\frac{1}{3}, 0.23604, 0.097285\right\}\\
\end{split}
\ees
with
\bes
\begin{split}
\alpha
&= \mathrm{Root}_1\left[-25957 + 163107 x - 78003 x^2 + 6561 x^3\right]\\
&= \frac{107}{27}-\frac{104\cdot2^{2/3} \left(1+i\sqrt{3}\right)}{27 \left(67+23 i \sqrt{7}\right)^{1/3}}-\frac{13\left(1-i\sqrt{3}\right) \left(67+23 i \sqrt{7}\right)^{1/3}}{27\cdot2^{2/3}}\\
&\doteq 0.17329.
\end{split}
\ees
In particular, $X$ is positive semidefinite with rank equal to $4$. Similarly, the linear independence of~\eqref{eq:extreme_unital_channel} can be verified explicitly.

\bibliographystyle{unsrt}
\bibliography{unital}

\end{document}